\documentstyle[tighten,floats,aps,epsf]{revtex}

\begin{document}
\title
{Structure of ground and excited states of $^{12}$C}

\author{Y. Kanada-En'yo}

\address{Yukawa Institute for Theoretical Physics, Kyoto University,\\
Kyoto 606-8502, Japan}

\maketitle
\begin{abstract}
  We studied the ground and excited states of $^{12}$C based on variational
calculations after spin-parity projection in a framework of antisymmetrized
molecular dynamics(AMD). The calculations systematically reproduce 
various experimental data. It was found that the 
sub-shell closure and $SU(3)$-limit $3\alpha$ cluster components are 
contained in the ground state, 
while various $3\alpha$ cluster structures develop in the excited states.
We discussed effects of $\alpha$ breaking and show the importance of 
coexistence of the cluster and shell-model-like aspects.

\end{abstract}

\noindent
%PACS numbers: 21.60.-n, 02.70.Ns, 21.10.Ky, 27.20.+n

\section{Introduction}

The formation of mean-field is an important aspect in nuclear systems. 
It results in the shell-model-like
nature of nuclear structure.
On the other hand, in a region of light nuclei, it is well known that 
clustering is another essential feature. 
These two different kinds of nature, cluster and 
shell-model-like aspects, often coexist, compete and interplay 
between each other.
The coexistence of two aspects provides a variety of structure in
nuclear systems, and can be a key to reveal properties of light nuclei. 

$^{12}$C is one of the typical examples where the 
cluster and shell-model-like aspects coexist.
Since $^{12}$C is the double sub-shell closure of the $p_{3/2}$ shell
in the $j$-$j$ coupling picture, 
the ground state should contain such the shell-model-like nature.
On the other hand, 
$3\alpha$-cluster structure is also favored in $^{12}$C and is considered
to develop in excited states.
Shell model calculations succeeded to reproduce experimental data 
of many levels in light nuclear region(for example, \cite{SHELL1,SHELL2}). 
However, a number of states in light nuclei have been left unsolved in those
shell model studies because
it is difficult for the shell models to describe well-developed
cluster states in general. In fact, 
even the large basis shell model calculations
failed to describe such the excited states as 
the $0^+_2$ state at $E_x=7.65$ MeV and 
the $3^-_1$ state at $E_x=9.64$ MeV near the $3\alpha$ 
threshold energy\cite{navratil03}.
These states have been considered to be well-developed cluster states, and
their structures 
have been studied with $3\alpha$ cluster models for 
a long time \cite{RGM,OCM,GCM,Fujiwara80,descouvemont87}.
Recently, Tohsaki {\em et al.} proposed a new interpretation of 
the $0^+_2$ state as a dilute cluster gas state of 
weakly interacting 3 $\alpha$ particles\cite{Tohsaki01}, and discussed it  
in the relation with the Bose Einstein Condensed(BEC) phase 
in dilute nuclear matter\cite{Ropke98}. 
Moreover, there are many experimental reports on 
the resonances in the excitaion energy region around
$E_x=10$ MeV recently 
\cite{john03,fynbo03,itoh04,diget05}.
Especially, the experimentally suggested 
$0^+$ and $2^+$ states in the
broad resonances at $E_x\sim 10$ MeV
attract an interest because these states are the 
candidates of the well-developed cluster states. 
Although the structure of these 
resonances\cite{kurokawa04,kurokawa05,funaki05,funaki06} 
has been theoretically 
studied based on the $3\alpha$ cluster model approaches, their properties 
have not been clarified yet.
The cluster model approaches are useful to describe the details of 
inter-cluster motion in the $3\alpha$ system, however, 
they failed to reproduce such the data
as the excitation energy of the $2^+_1$ and the strengths of 
$\beta$ decays, because these properties are sensitive to 
such the shell-model-like features as 
the sub-shell closure and the $\alpha$ breaking of $^{12}$C but the
$3\alpha$-cluster models is not suitable for describing  
the $j$-$j$ coupling features nor dissociation of the $\alpha$ clusters.

These facts indicate that the cluster and shell-model-like
features certainly coexist in the $^{12}$C system. 
It is expected that the shell-model-like aspects
in the low-lying states may have some influence on the excited cluster
states through the orthogonality and mixing with the lower states.
Therefore, it is important to take into account both aspects
in the systematic study of $^{12}$C.
However, there are a few theoretical studies of $^{12}$C where 
both the cluster and the shell-model-like features are 
taken into account. The method of antisymmetrized molecular
dynamics(AMD) has been proved to be a powerful approach to 
describe the cluster and shell-model-like features in general nuclei.
Since all the single-nucleon wave functions are treated as 
independent Gaussian wave packets in this method, 
the existence of clusters are not assumed
in the model. Instead, if the system favors a cluster structure, 
the cluster state
automatically appears in the energy variation in this framework.
For the excited states, the method of variation after spin-parity
projection(VAP) in the framework of AMD is a useful
approach\cite{ENYO-c12,ENYO-be10,ENYO-be12}.
The author has performed the first VAP calculations of AMD 
and briefly reported its application to $^{12}$C\cite{ENYO-c12}.
Itagaki {\em et al.} also
have investigated the cluster and shell competition in the $^{12}$C system 
while incorporating the dissociation of  
$\alpha$ clusters\cite{Itagaki04}. However, the work was 
limited to the low-lying states.
Neff {\em et al.} have done the VAP calculations within the
fermionic molecular dynamics(FMD)\cite{Neff-c12}, 
which is similar approach to the AMD. Although the FMD wave function 
is suitable for describing both the cluster
and shell-model-like structures in principle as well as the AMD, 
the existence of 3 $\alpha$-clusters 
were {\it apriori} assumed in the practical application to 
$^{12}$C in\cite{Neff-c12}.

In the present work, we studied the structure of the ground and excited
states of $^{12}$C with the method of the VAP in the AMD framework.
The method is almost the same as that applied 
in our previous work\cite{ENYO-c12}. We calculated various quantities such
as radii, transition strengths and densities, and compared them with 
the experimental data and other theoretical calculations. The structures
of the ground and excited states were analyzed while 
focusing on the cluster structure and the 
dissociation of the $\alpha$ clusters. We also estimated the partial 
decay widths of the resonances with the method of reduced with amplitudes.

This paper is organized as follows.
In \ref{sec:formulation}, we give a brief explanation of the formulation
of the present work. The adopted effective nuclear interactions are described
in \ref{sec:interactions}.
In \ref{sec:results}, we show the calculated results concerning such
observables as the energy levels, radii and $\beta$ decays as well as the
$E0$ and $E2$ transitions compared with the experimental data.
In an analysis of the obtained wave functions in \ref{sec:discuss}), 
the shell-model-like and cluster features 
are discussed.
In the last section(\ref{sec:summary}), we give a summary.

\section{Formulation}
 \label{sec:formulation}

The detailed formulation of the AMD method 
for nuclear structure studies is described in 
\cite{ENYO-c12,ENYO-be10,ENYObc,ENYOsup,AMDrev}.
In particular, the formulation of the present calculations is basically 
the same as that described in \cite{ENYO-c12,ENYO-be10,ENYO-be12}.

An AMD wave function is a Slater determinant of Gaussian wave packets;
\begin{equation}
 \Phi_{\rm AMD}({\bf Z}) = \frac{1}{\sqrt{A!}} {\cal{A}} \{
  \varphi_1,\varphi_2,...,\varphi_A \},
\end{equation}
where ${\cal A}$ is the antisymmetrizer and 
$A$ is the mass number. The $i$-th single-particle 
wave function is written by a product of
spatial($\phi$), intrinsic spin($\chi$) and isospin($\tau$) 
wave functions as,
\begin{eqnarray}
 \varphi_i&=& \phi_{{\bf X}_i}\chi_i\tau_i,\\
 \phi_{{\bf X}_i}({\bf r}_j) &\propto& 
\exp\bigl\{-\nu({\bf r}_j-\frac{{\bf X}_i}{\sqrt{\nu}})^2\bigr\},
\label{eq:spatial}\\
 \chi_i &=& (\frac{1}{2}+\xi_i)\chi_{\uparrow}
 + (\frac{1}{2}-\xi_i)\chi_{\downarrow}.
\end{eqnarray}
The spatial part is represented by 
complex variational parameters, ${\rm X}_{1i}$, ${\rm X}_{2i}$, 
${\rm X}_{3i}$, which indicate the center of the Gaussian wave packet.
The orientation of the intrinsic spin is expressed by
a complex variational parameter $\xi_{i}$, and the isospin
function is fixed to be up(proton) or down(neutron). 
We used a common value of the width parameter $\nu$, which 
is chosen to be the optimum value of each nucleus.
Accordingly, an AMD wave function
is expressed by a set of variational parameters, ${\bf Z}\equiv 
\{{\bf X}_1,{\bf X}_2,\cdots, {\bf X}_A,\xi_1,\xi_2,\cdots,\xi_A \}$.

We performed the energy variation after spin-parity 
projection(VAP) in the AMD model space. 
In order to obtain the wave function for the lowest $J^\pi$ state,
we varied the parameters ${\bf X}_i$ and $\xi_{i}$($i=1\sim A$) to
minimize the energy expectation value of the Hamiltonian,
$\langle \Phi|H|\Phi\rangle/\langle \Phi|\Phi\rangle$,
for the spin-parity projected AMD wave function;
$\Phi=P^{J\pi}_{MK'}\Phi_{\rm AMD}({\bf Z})$.
$P^{J\pi}_{MK'}$ is the spin-parity projection operator.
By the VAP calculation for the $J^\pi$ eigen state
with an appropriate $K'$, we first 
obtained a set of parameters ${\bf Z}={\bf Z}^{J\pi}_1$ for the
lowest $J^\pi$ state
which is expressed by $P^{J\pi}_{MK'}\Phi_{\rm AMD}({\bf Z}^{J\pi}_1)$. 
In order to search for the parameters ${\bf Z}$ for the 
higher $J^\pi$ state, the VAP is performed in the orthgonal space
to the lower states.
The parameters ${\bf Z}^{J\pi}_n$ 
for the $n$th $J^\pi$ state are provided by varying ${\bf Z}$ 
so as to minimize the energy of the wave function; 
\begin{equation}
|\Phi\rangle =|P^{J\pi}_{MK'}\Phi_{\rm AMD}({\bf Z})\rangle
-\sum^{n-1}_{k=1}
|P^{J\pi}_{MK'}\Phi_{\rm AMD}({\bf Z}^{J\pi}_k)\rangle 
\frac{
\langle P^{J\pi}_{MK'}\Phi_{\rm AMD}({\bf Z}^{J\pi}_k)
||P^{J\pi}_{MK'}\Phi_{\rm AMD}({\bf Z})\rangle}
{\langle P^{J\pi}_{MK'}\Phi_{\rm AMD}({\bf Z}^{J\pi}_k)|P^{J\pi}_{MK'}\Phi_{\rm AMD}({\bf Z}^{J\pi}_k)\rangle},
\end{equation}
which is orthogonalized to the lower states. 

After the VAP calculation of the $J^\pi_n$ states for various
$J$, $n$ and $\pi=\pm$,
we obtained the optimum  intrinsic states,
$\Phi_{\rm AMD}({\bf Z}^{J\pi}_n)$, 
which approximately describe the corresponding $J^\pi_n$ states. 
In order to improve the wave functions, we superposed all the 
obtained AMD wave functions. 
Namely, we determined the final wave functions for the $J^\pi_n$ states
by simultaneously diagonalizing the Hamiltonian matrix, 
$\langle P^{J\pi}_{MK'} \Phi_{\rm AMD}({\bf Z}^{J_i \pi_i}_{n_i})
|H|P^{J\pi}_{MK''} \Phi_{\rm AMD}({\bf Z}^{J_j \pi_j}_{n_j})\rangle$,
and the norm matrix,
$\langle P^{J\pi}_{MK'} \Phi_{\rm AMD}({\bf Z}^{J_i\pi_i}_{n_i})
|P^{J\pi}_{MK''} \Phi_{\rm AMD}({\bf Z}^{J_j\pi_j}_{n_j})\rangle$,
with respect to ($i,j$) for 
all of the obtained intrinsic states and 
($K', K''$).
Consequently, the 
$J^\pi_n$ state are written as,
\begin{equation}\label{eq:diago}
|J^\pi_n\rangle=\sum_{i,K} c(J^\pi_n,i,K) 
|P^{J\pi}_{MK}\Phi_{\rm AMD}({\bf Z}^{J_i\pi_i}_{k_i})\rangle,
\end{equation}
where the coefficients $c(J^\pi_n,i,K)$ are determined by the 
diagonalization of the Hamiltonian and norm matrices.
In comparison of theoretical values with the experimental data, 
we calculated the expectation values for the corresponding operators
by the $|J^\pi_n\rangle$.

\section{Interactions and parameters} 
\label{sec:interactions}

The effective nuclear interaction adopted in the present work
consists of the central force, the
spin-orbit force and the Coulomb force.
We adopted MV1 force \cite{TOHSAKI} as the central force,
which contains a zero-range three-body force
in addition to the two-body interaction.
Concerning the spin-orbit force, the same form of the two-range Gaussian 
as the G3RS force \cite{LS}
is adopted.
The adopted interaction parameters are the same as those 
used in the previous work\cite{ENYO-c12}. Namely, 
the Bartlett, Heisenberg and Majorana parameters in the MV1 force
are chosen to be $b=h=0$ and $m=0.62$, and the 
strengths of the spin-orbit force are taken to be 
$u_{I}=-u_{II}=3000$ MeV.

We used the width parameter $\nu=0.19$ fm$^{-2}$ 
for the single-particle Gaussian wave packets 
in Eq.~\ref{eq:spatial} so as to minimize the energy of the
lowest $0^+$ state.

 In the previous work \cite{ENYO-c12}, we situated an artificial
barrier potential in the procedure of the energy variation 
to confine nucleons within the inner region
in the same way as Ref.\cite{ENYO-be10}.
Since we are interested in the dilute cluster states, we
performed the energy variation without the barrier potential.
We also did the variation with the barrier potential and combined
the obtained wave functions as the base wave functions 
in the diagonalization of Hamiltonian and norm matrices.

\section{Results}
\label{sec:results}

The wave functions of the ground and excited states of $^{12}$C
were calculated based on the VAP calculations in the framework of 
AMD. In this section, we show theoretical results and compare them
with experimental data. 

  The wave functions for the lowest $J^\pi$ states were obtained by
the VAP calculation of $P^{J\pi}_{MK'}\Phi_{\rm AMD}$ by using
$(J^\pi,K')$=
$(0^+,0)$, $(2^+,0)$, $(4^+,0)$, $(6^+,0)$, 
$(1^-,1)$, $(2^-,1)$, $(3^-,3)$, $(4^-,3)$, $(5^-,3)$,
$(1^+,0)$. 
After obtaining the lowest states($J^\pi_1$), we calculated the 
second and third $J^\pi$ states ($0^+_2$, $0^+_3$, 
$2^+_2$, $2^+_3$, $4^+_2$)
with the VAP calculation in the model space orthogonal 
to the obtained lower $J^\pi$ states.

  Those VAP calculations were performed without the artificial
barrier potential. As shown later, thus obtained 
intrinsic wave functions for the $0^+_3$, $2^+_2$, $2^+_3$, 
$4^+_2$, $6^+_1$, $1^-_1$, $2^-_1$ states are not compact states but
spatially expanded states where $\alpha$ particles distribute 
far away from the center. For these states, we also performed 
the VAP calculations with the artificial barrier potential
\cite{ENYO-be10} in the same way as the previous work 
\cite{ENYO-c12}
to keep the particles in an intermediate distance region.

The obtained AMD wave functions($\Phi_{\rm AMD}({\bf Z}^{J\pi}_n)$) 
are considered to approximately 
describe the intrinsic states of the corresponding $J^\pi_n$ states.
The final wave functions of the $J^\pi_n$ states were determined by 
superposing the spin-parity eigen states projected from these obtained
AMD wave functions so as to simultaneously diagonalize the Hamiltonian
and the norm matrices. 
For the $1^+$ state,
we found that the obtained wave function,
$P^{1+}_{M1}\Phi_{\rm AMD}({\bf Z}^{1+})$, 
contains a significant isospin breaking component. Therefore, we added the 
mirror wave function 
$P_{p\leftrightarrow n}P^{1+}_{M1}\Phi_{\rm AMD}({\bf Z}^{1+})$ 
of the original one to the set of base AMD wave functions in the 
diagonalization.
Here $P_{p\leftrightarrow n}$ is the proton-neutron exchange operator.
As a result, the number $i$ of the superposed AMD wave functions 
$P^{J\pi}_{MK'} \Phi_{\rm AMD}({\bf Z}^{J_i\pi_i}_{n_i})$
is 23 in the present calculations.

\subsection{Energies}
  The theoretical binding energy of $^{12}$C is 88.0 MeV.
Although the calculation slightly underestimates
the experimental value 92.16 MeV, it can be improved by changing
the Majorana parameter $m$ of the interaction.

\begin{figure}%----------------------------------------
\noindent
\epsfxsize=0.49\textwidth
\centerline{\epsffile{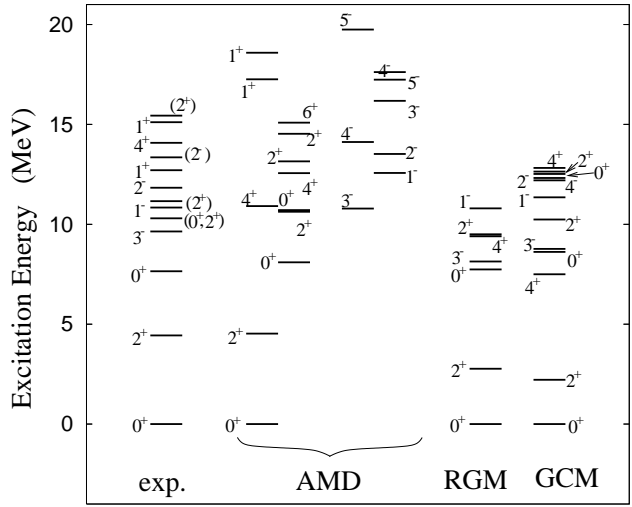}}
\caption{\label{fig:sped}
Energy levels of the ground and excited states of $^{12}$C.
The theoretical levels of the $3\alpha$RGM and
$3\alpha$GCM are the results of \protect\cite{RGM} and \protect\cite{GCM},
respectively.}
\end{figure}

 In Fig.~\ref{fig:sped}, the energy levels of the present results(AMD)
are shown with the experimental data and 
the other calculations of $3\alpha$ cluster models;
the resonating group method(RGM)\cite{RGM} 
with the Volkov No.2 force and the
generator coordinate method(GCM)\cite{GCM} 
with the Volkov No.1 force. 
Recently, further theoretical works
were performed based on $3\alpha$ cluster models, 
where resonant 3$\alpha$ states were treated in more 
details\cite{kurokawa04,kurokawa05,funaki05,funaki06}.
The energy spectra obtained in these works are almost consistent
with either of the $3\alpha$RGM\cite{RGM} and 
$3\alpha$GCM\cite{GCM} results shown in Fig.~\ref{fig:sped}, except for the
new broad $0+$ state predicted in \cite{kurokawa05}.

 In the ground band, the AMD calculation reproduces 
well the level spacing between 
$0^+_1$ and $2^+_1$, while all the $3\alpha$ calculations underestimate
it. The success of the AMD is owing to the energy gain of the 
spin-orbit force in the $0^+_1$ with the breaking of the 3$\alpha$ clustering.
In other words, this large level spacing is one of the characteristics
of the $p_{3/2}$ sub-shell closure, which
can not be described within the $3\alpha$ cluster models.

For the positive parity states with natural spins
above the 3$\alpha$ threshold energy(7.272 MeV), the 
$0^+_2$, $0^+_3$, $2^+_2$, $2^+_3$,  $4^+_1$, $4^+_2$ and $6^+_1$ states 
are obtained in the AMD results.
The level structure of these states is similar to that of the 
$3\alpha$GCM calculations\cite{GCM}. 
Since the highly excited states, $0^+_3$, $2^+_2$ ,$2^+_3$ and $6^+_1$, 
have the dominant component
with an $\alpha$-particle far from the other two $\alpha$'s,
the stability of these states should be analyzed
by treating the boundary condition of the resonant states carefully. 

The level spectra of the positive parity states around $E_x=10\sim 11$ MeV
and their properties have not been clarified
experimentally and theoretically. In the AMD results, the $0^+_3$
state appears at 3 MeV higher energy than the $0^+_2$ state.
The $0^+_3$ state was predicted also in the $3\alpha$GCM 
calculation\cite{GCM}, 
while Kurokawa {\em et al.} proposed a new broad $0^+$ state below the
$0^+$ in the recent work with the method of 
analytic continuation in the coupling constant 
combined with the complex scaling method(ACCC+CSM)\cite{kurokawa05}.
As discussed later, the $0^+_3$ in the present result 
corresponds to the $0^+_3$ state of the $3\alpha$GCM calculation.
The detailed assignment of the excited states is discussed in the next
section.

We obtain the $1^+_1$ and $1^+_2$ states. These states correspond to the 
$1^+(T=0)$ at 12.7 MeV and the $1^+(T=1)$ at 15.1 MeV. 
Since the spin-parity $1^+$ is unnatural in a
$3\alpha$ system, it is difficult to describe these states within 
$3\alpha$ cluster models. In fact, the present $1^+$ states 
have the non-$3\alpha$ component.
The excitation energies of the $1^+$ states 
are overestimated in the present results.
They can be improved by tuning the strength of Bartlett 
and Heisenberg terms. For example, by using the interaction parameters, 
$b=h=0.2$ and $m=0.62$, we obtained the excitation energy 
$E_x(1^+_1)=13.9$ MeV.

In the negative parity states, we obtained 
\{$3^-_1$, $4^-_1$, $5^-_2$\} and 
\{$1^-_1$, $2^-_1$, $3^-_2$, $4^-_2$, $5^-_1$\}, which 
construct the rotational bands, 
$K=3^-$ and $K=1^-$, respectively.
These bands are consistent with the 3$\alpha$GMC calculations.

\subsection{Radii}

The theoretical values of the root mean square radius($R_{m}$) 
of mass distributions
are shown in Table \ref{tab:radii}. The calculated radius
of the ground state is $R_m=2.53$ fm, which is slightly larger than
other $3\alpha$ calculations and is large compared with
the observed value $R_m=2.32-2.33$ fm deduced from the charge radius.
The radii($R_m$) of the excited states 
in the present results seem to be qualitatively consistent
with the those of the $3\alpha$GCM calculations.

Recently, the structure of the $0^+_2$ state attracts a great interest,
because Tohsaki {\em et al.}\cite{Tohsaki01} proposed a new interpretation 
of the $0^+_2$ as a dilute $\alpha$-cluster gas,
where weakly interacting 3 $\alpha$'s form an $\alpha$ condensate state.
In their work, the $0^+_2$ has the extremely large radius 
because of its dilute $3\alpha$ gas-like structure.
Although the radius of the $0^+_2$ is somewhat
smaller in the present result than the theoretical values of 
the $3\alpha$ calculations, it is still remarkably
large compared with that of the ground state. In the present 
framework, the long tail of the inter-cluster 
wave functions at a dilute density 
may be underestimated because the base AMD wave functions is 
limited to only 23 Slater determinants obtained by the VAP.
We think that the dissociation of the $\alpha$ clusters can be 
another reason for the smaller radius of the $0^+_2$ than 
the $3\alpha$ calculations.
One of the origins of the dilute character of the $0^+_2$ state is the 
orthogonality to the ground state. In the $3\alpha$ cluster model,
the ground state has the compact $3\alpha$ wave function,
the wave function of the $0^+_2$ state tends to avoid overlapping with 
the compact $3\alpha$ state to keep the othogonality. However, in the 
present results, the ground state
contains the $\alpha$ breaking component
as naturally expected in the $p_{3/2}$ sub-shell closed nucleus. 
Due to the non-$3\alpha$ component, the compact $3\alpha$ wave function
is partially allowed to mix into the $0^+_2$ state, and it may reduce the
size of the $0^+_2$ state. 

\begin{figure}%----------------------------------------
\noindent
\epsfxsize=0.49\textwidth
\centerline{\epsffile{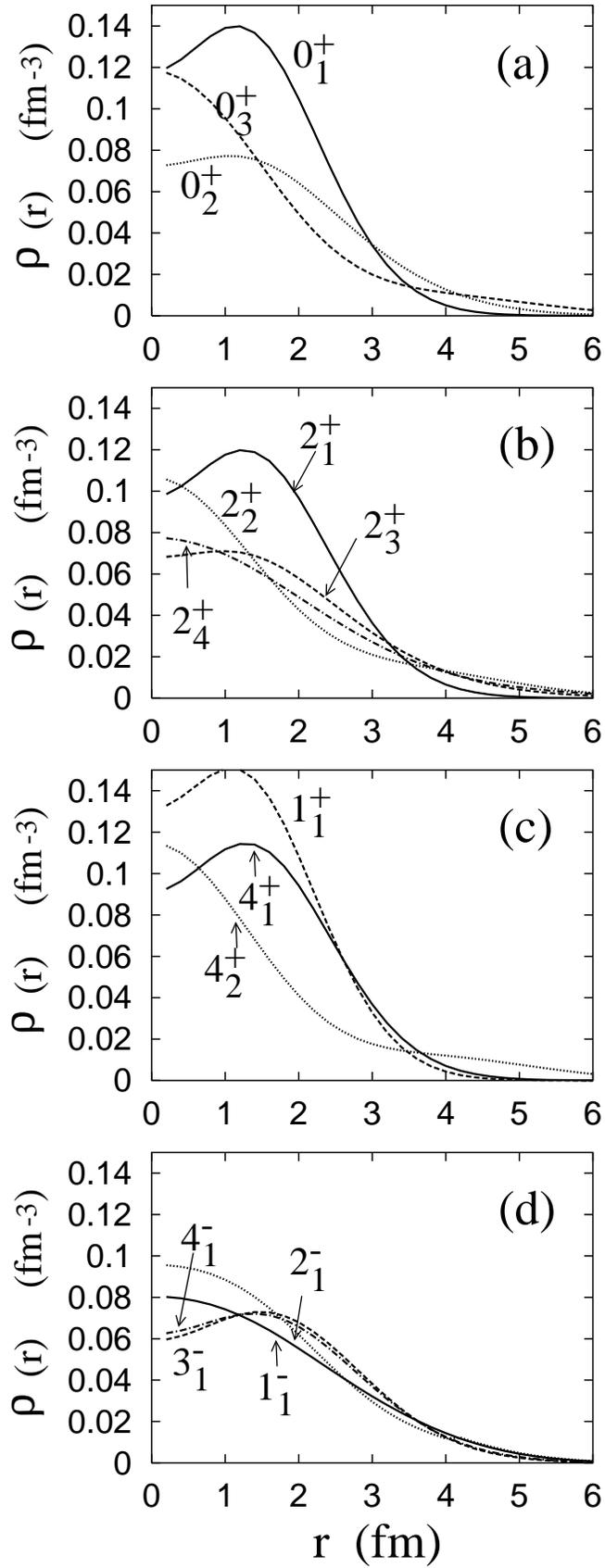}}
\caption{\label{fig:dens}
Calculated matter density plotted as a function of 
radius $r$.
}
\end{figure}

In Fig.~\ref{fig:dens}, the matter density distributions are shown.  
The $0^+_1$, $2^+_1$ and $4^+_1$ in the ground band have
compact density distributions. The density distributions of the
$1^+$ states are also compact and its radius is small
because of the shell-model-like structure with
the dominant $0\hbar\omega$ components as well as the ground state. 
In the $0^+_2$ state, 
the density is suppressed in the small $r$ region and spreads to
 the outer region.
The shapes of the densities in the $0^+_3$, $2^+_2$ and $4^+_2$ are
similar to each other. They have long tails due to the $3\alpha$ 
linear-like structure. 
In the $3^-_1$ and $4^-_1$ states, the density at the center $r=0$ fm 
is small, and the positions of the maximum density
are large as $r\sim 1.5$ fm compared with those in other states
because of a developed triangle $3\alpha$ cluster 
structure.

\begin{table}
\caption{\label{tab:radii} Root mean square radius for 
mass distributions. The theoretical values of the $3\alpha$-cluster 
calculations\protect\cite{RGM,GCM,funaki05} are also listed.
The values for the $3\alpha$ condensate wave functions in 
Ref.\protect\cite{funaki05} are
those obtained with the Volkov No.2 force. The mass 
radii($R_m$) for the $3\alpha$RGM calculations\protect\cite{RGM} 
are deduced from the charge radii($R_c$) 
with the relation $R_c^2=R_m^2-R_p^2$, where
$R_p$ is the proton charge radius 0.813 fm. 
The observed charge radius of the ground state 
is $R_c=2.46-2.47$ fm\protect\cite{nucldata}, 
which corresponds to 
$R_m=2.32-2.33$ fm.}
\begin{center}
\begin{tabular}{ccccc}
$J^\pi$ &  \multicolumn{4}{c}{mass radius (fm)}\\ 
  & present(AMD) &  $3\alpha$GCM\cite{GCM} & $3\alpha$RGM\cite{RGM} & 
Funaki {\em et al.}\cite{funaki05}. 
\\
\hline
$0^+_1$ & 2.53  & 2.40 & 2.40 & 2.40 \\
$0^+_2$ & 3.27 & 3.40 & 3.47 &  3.83\\
$0^+_3$ & 3.98 & 3.52  &  & \\
$1^+_1$ & 2.47 & & & \\
$1^+_2$ & 2.47 & & & \\
$2^+_1$ & 2.66 & 2.36  & 2.38 & 2.38\\
$2^+_2$ & 3.99 & 3.52  & 4.0 & \\
$2^+_3$ & 3.50 & 3.34  &  & \\
$2^+_4$ & 3.86 &  &  & \\
$4^+_1$ & 2.71 & 2.29  & 2.31 & 2.31\\
$4^+_2$ & 4.16 & 3.64  &  & \\
$1^-_1$ & 3.42 & 3.29  & 3.36 & \\
$2^-_1$ & 3.49 & 3.32  &  & \\
$3^-_1$ & 3.13 & 2.83  & 2.76 &  \\
$3^-_2$ & 3.56 &   & &  \\
$4^-_1$ & 3.19 & 2.87  &  & \\
$4^-_2$ & 3.53 &   &  &  \\
\end{tabular}
\end{center}
\end{table}

\subsection{$E2$ and $E0$ transitions}

The $E2$ and $E0$ transition strengths are listed in Table \ref{tab:e2}.
The present $B(E2;2^+_1\rightarrow 0^+_1)$ value for the transitions
in the ground band well agrees with the experimental 
data and is consistent with
other theoretical calculations.
For the transition between the $2^+_1$ and the $0^+_2$, the 
$3\alpha$ cluster calculations\cite{GCM,RGM} 
usually underestimate the observed 
data by factor $2\sim 4$. In the present results, the 
$B(E2;2^+_1\rightarrow 0^+_2)$ value is larger than those of the
$3\alpha$ calculations. It is because 
the spatial extension of the $3\alpha$ distribution in the $0^+_2$ 
is not so remarkable as the $3\alpha$ calculations
due to the $\alpha$ breaking component of the ground state, 
and can have a significant transition overlap with
the $2^+_1$ state. It means that the $\alpha$ breaking component affect the
$B(E2;2^+_1\rightarrow 0^+_2)$ as argued in the work with
the $\alpha+4p+4n$ model by Itagaki {\em et al.}\cite{Itagaki04}.
Considering that the present result overestimates the 
observed $B(E2;2^+_1\rightarrow 0^+_2)$, the $\alpha$ breaking effect 
might be somewhat too strong.
In the $E2$ transitions from the highly excited states
such as the $2^+_2$, $0^+_3$, $4^+_2$, we obtained
remarkably large $B(E2)$ values,
$B(E2;2^+_2\rightarrow 0^+_2)=100$ e$^2$fm$^4$,
$B(E2;2^+_2\rightarrow 0^+_3)=310$ e$^2$fm$^4$
and $B(E2;4^+_2\rightarrow 2^+_2)=600$ e$^2$fm$^4$,
due to the large intrinsic deformation of the 
developed $3\alpha$ cluster states.
Since the $E2$ transitions $2^+_2\rightarrow 0^+_3$ 
and $4^+_2\rightarrow 2^+_2$ are especially strong, we consider 
that those are the intra-band transitions among the band members 
\{$0^+_3$, $2^+_2$ and $4^+_2$\}.
This assignment of the band members is also consistent 
with the similarity of the density form among
these three states(see Fig.~\ref{fig:dens}). The detailed discussion of 
the band structure is given in the next section. 

\begin{figure}%----------------------------------------
\noindent
\epsfxsize=0.49\textwidth
\centerline{\epsffile{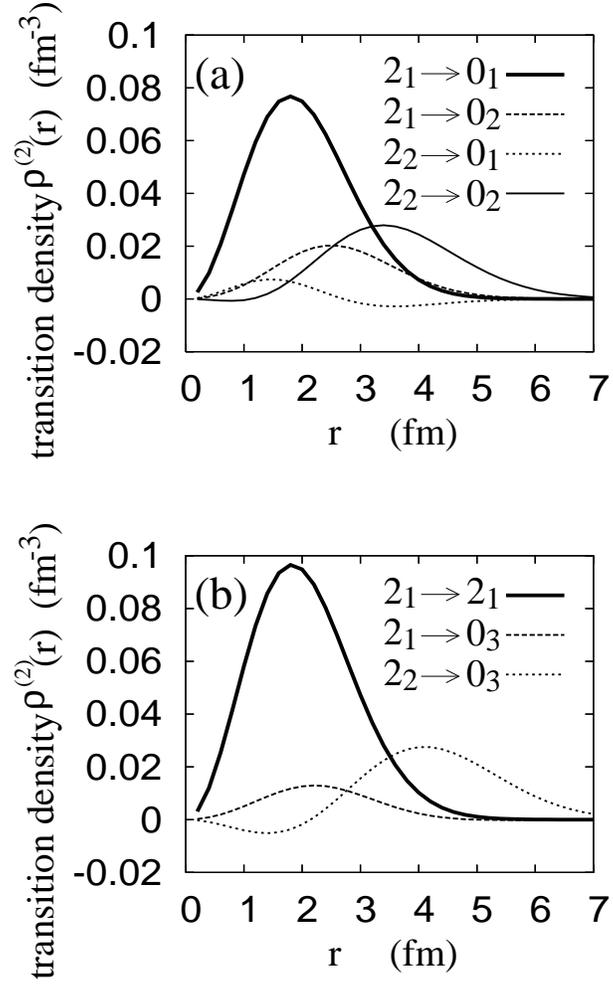}}
\caption{\label{fig:trans}
Calculated $E2$ transition densities $\rho^{(\lambda=2)}_{J_f,J_i}(r)$ between
the $0^+$ and $2^+$ states. }
\end{figure}

The $E2$ transition densities $\rho^{(\lambda=2)}_{J_f,J_i}(r)$ are shown  
in Fig.~\ref{fig:trans}. The density is normalized as,
\begin{equation}
B(E2;J_i\rightarrow J_f)=\frac{1}{2J_i+1}\int r^4 dr \rho^{(\lambda=2)}_{J_f,J_i}(r).
\end{equation}
The shapes of the transition densities 
for $2^+_1\rightarrow 0^+_1$, $2^+_2\rightarrow 0^+_1$, and 
$2^+_2\rightarrow 0^+_2$ in the 
present results are consistent with those of the $3\alpha$RGM 
calculations\cite{RGM}. 
The transition density for $2^+_1\rightarrow 0^+_2$ 
has a significant strength in the region $r =2\sim 3$ fm in the present results, 
while, in the $3\alpha$RGM case, it has a node 
which suppresses $B(E2)$.

\begin{figure}%----------------------------------------
\noindent
\epsfxsize=0.49\textwidth
\centerline{\epsffile{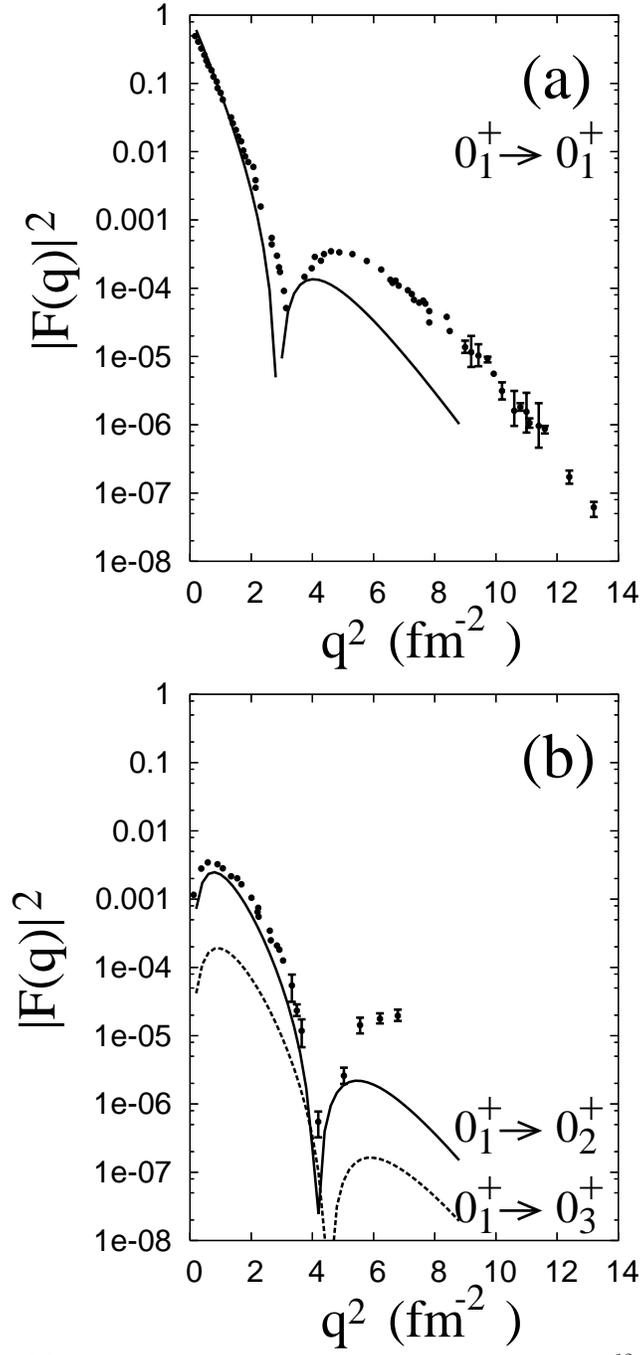}}
\caption{\label{fig:form}
Squared (a)elastic and (b)inelastic form factors 
of the electron scattering from $^{12}$C 
given by the calculated transition densities and by the 
observation\protect\cite{Sick70}.}
\end{figure}

The elastic form factor for the ground state 
is shown in Fig.~\ref{fig:form} (a). The dip position shifts toward the small
$q$ region compared with the observed data, because the present calculation
overestimates the radius of the ground state 
as shown in Table \ref{tab:radii}.
The inelastic form factors for the electric monopole transitions,
$0^+_1 \rightarrow 0^+_2$ and $0^+_1 \rightarrow 0^+_3$ 
are shown in Fig.~\ref{fig:form}(b). The first peak of the form factor 
for $0^+_1 \rightarrow 0^+_2$ 
well agrees with the observed data for the 
inelastic scattering into the $0^+_2$ state at 7.65 MeV. 
The second peak at the high $q$ region is smaller than the 
observed data because the calculation underestimates the 
observed elastic form factor of the ground state 
in this region.
Recently, Funaki {\em et al.} discussed the dependence of the 
inelastic form factor on the nuclear size of the $0^+_2$
within the 3$\alpha$ cluster model\cite{funaki-nucl06}.
They argued that the maximum value of the form factor
depends on the size.
If the $0^+_2$ state with the 3$\alpha$ structure has a small size,
the inelastic form factor becomes large because of the large
transition overlap with the compact ground state.
In their analysis the observed form factor is well reproduced 
by the $\alpha$ condensate wave function with the radius $R_m=3.8$ fm.
Also in the $3\alpha$GCM and $3\alpha$RGM calculations,
the form factor is well reproduced by the $0^+_2$ with $R_m=3.5$ fm.
On the other hand, although the size of the $0^+_2$ is as small as 3.3 fm
in the present results,
the maximum peak height of the calculations is in good agreement with 
that of the observed inelastic form factor. 
In the present case, 
the $\alpha$ breaking component in the ground state
suppresses the transition into the $3\alpha$ cluster state, 
because the non-$3\alpha$ component 
has a small transition overlap with the $3\alpha$ cluster state.
It means that the absolute value of the inelastic form factor 
is not sensitive only to the size of the $0^+_2$ state, 
but also depends on the $\alpha$ breaking component in the ground state.

The calculated form factor for the $0^+_1 \rightarrow 0^+_3$ 
has a similar $q$ dependence to that of the $0^+_1 \rightarrow 0^+_2$.
Its magnitude is one order smaller that for the 
$0^+_1 \rightarrow 0^+_2$ in the 
present results. It corresponds to the 3 times smaller 
transition matrix $M(E0;0^+_1\rightarrow 0^+_3)=2.0$ fm$^2$ 
than the $M(E0;0^+_1\rightarrow 0^+_2)=6.0$ fm$^2$.
The isoscalar monopole strengths to the $0^+_2$ and the $0^+_3$
were  studied by inelastic $^6$Li
and $\alpha$ scattering
\cite{john03,eyrich87}.
From the isoscalar energy weighted sum rule strength in those 
studies, the ratio 
$B(E0;0^+_1\rightarrow 0^+_3)/B(E0;0^+_1\rightarrow 0^+_2)$
is deduced to be $0.3\sim 1$ by assuming the mirror symmetry. 
The present form factor for the $0^+_1 \rightarrow 0^+_3$
is smaller than these experimental values. 
In the $3\alpha$GCM calculations\cite{GCM}, 
the inelastic form factor for $0^+_1 \rightarrow 0^+_3$
is as large as that for $0^+_1 \rightarrow 0^+_2$
and is consistent with the observed data. 
In order to study the detailed structure of the $0^+_3$ state, 
more precise experimental data are required.

\begin{table}
\caption{\label{tab:e2} Strengths of $E2$ and $E0$ transitions.
The experimental data are taken from Ref.\protect\cite{isotopes}.
The theoretical results of $3\alpha$GCM\protect\cite{GCM}, 
$3\alpha$RGM\protect\cite{RGM} 
and $\alpha+4p+4b$ calculations\protect\cite{Itagaki04} are 
also listed.}
\begin{center}
\begin{tabular}{cccccccc}
transitions &  present(AMD) &  
$3\alpha$GCM\protect\cite{GCM} & $3\alpha$RGM\protect\cite{RGM} & 
$\alpha+4p+4n$\protect\cite{Itagaki04}& exp. \\
\hline
$B(E2;2^+_1\rightarrow 0^+_1)$ & 8.5  & 8.0 & 9.3 & 7.1  & 7.6$\pm$ 0.4 $e^2$fm$^4$ \\
$B(E2;2^+_1\rightarrow 0^+_2)$ & 5.1  & 0.7 & 1.1 & 2.8  & 2.6$\pm 0.4$ $e^2$fm$^4$ \\
$B(E2;2^+_2\rightarrow 0^+_2)$ & 100 &  &  & &\\
$B(E2;2^+_2\rightarrow 0^+_3)$ & 310 &  &  & &\\
$B(E2;2^+_3\rightarrow 0^+_2)$ & 6.4 &  &  & &\\
$B(E2;2^+_3\rightarrow 0^+_3)$ & 76 &  &  & &\\
$B(E2;4^+_1\rightarrow 2^+_1)$ & 16 &  &  & &\\
$B(E2;4^+_1\rightarrow 2^+_2)$ & 7.5 &  &  & &\\
$B(E2;4^+_2\rightarrow 2^+_2)$ & 600 &  &  & &\\
$B(E2;4^+_2\rightarrow 2^+_3)$ & 74   &  &  & &\\
$M(E0;0^+_1\rightarrow 0^+_2)$ & 6.7  & 6.6  & 6.7  &  &  5.4$\pm 0.2$ fm$^2$ \\
$M(E0;0^+_2\rightarrow 0^+_3)$ & 2.0  &      &     & 
\end{tabular}
\end{center}
\end{table}

\subsection{GT and $M1$ transitions}

The $\beta$ decay strengths from neighboring nuclei, 
$^{12}$B and $^{12}$N to the ground and excited states of $^{12}$C 
are experimentally known. 
The observed values are of great help to investigate the 
structure of $^{12}$C as well as the electric transition strengths. 
However, within the $3\alpha$ cluster models, it is difficult to discuss 
the $\beta$ transitions  because
Gamov-Teller(GT) transition matrix for the ideal $\alpha$ cluster
of the $(0s)^4$ configuration is exactly zero due to the Pauli principle.
In other words, GT transitions 
to the $3\alpha$ cluster component are forbidden, but the transitions are
caused by the breaking of the $\alpha$ clusters. As a result,
the $\beta$ decay strengths are sensitive to the
$\alpha$ breaking component.

In Table \ref{tab:beta}, 
the calculated GT transition strengths for the $\beta^+$-decay
from $^{12}$N are
shown and compared with the 
experimental data. The ground state($J^\pi=1^+$) of the parent nucleus $^{12}$N
is obtained with a variational calculation after spin-parity projection to
the $1^+$ with the width parameter $\nu=0.19$ fm$^{-2}$.
The calculated strengths are in good agreement with the experimental data.
The strong $\beta$ decay to the ground state is because of the
rather large mixing of the $\alpha$ breaking component, which reflects 
the shell-model-like feature of the $j$-$j$ coupling.
The GT transitions to the $0^+_2$ and $0^+_3$ states have significant
strengths though these states are dominated by the 
developed $3\alpha$ cluster states. 
The strengths of the transitions to these excited $0^+$ states come from 
the mixing of the wave functions in the diagonalization.
Before the diagonalization, 
it was found that
the GT transitions are weak for 
the original single AMD wave functions,
$P^{0+}_{MK}\Phi_{\rm AMD}({\bf Z}^{0+}_{2,3})$ for the $0^+_2$ and $0^+_3$,
where the $3\alpha$ cluster structure is developed.
After the diagonalization, the $P^{0+}_{MK}\Phi_{\rm AMD}({\bf Z}^{0+}_1)$,
$P^{0+}_{MK}\Phi_{\rm AMD}({\bf Z}^{0+}_2)$ and
$P^{0+}_{MK}\Phi_{\rm AMD}({\bf Z}^{0+}_3)$ mix in the superposed 
wave functions, and the final wave functions $|0^+_2\rangle$
and $|0^+_3\rangle$ contain small fraction of the 
$P^{0+}_{MK}\Phi_{\rm AMD}({\bf Z}^{0+}_1)$.
As a result, the $\alpha$ breaking component, which is originally 
included in the ground state wave function
$P^{0+}_{MK}\Phi_{\rm AMD}({\bf Z}^{0+}_1)$, 
is contained in the $|0^+_2\rangle$ and the $|0^+_3\rangle$ 
through the superposition of the wave functions. This is the origin of the
significant GT transitions to the $0^+_2$ and the $0^+_3$.
The GT transition to the $2^+_1$ is weaker than that to the $0^+_1$.
It is reasonable because the $3\alpha$ core structure enhances in the
$2^+_1$ compared with the $0^+_1$ as shown in the next 
section. 
The strong GT transitions to the $1^+_1(T=0)$ state at 12.71 MeV is 
described well by the present calculations
reflecting the intrinsic spin excitation in this state.

The calculated $M1$ transition strengths are shown in Table \ref{tab:bm1}
compared with the experimental data.
The results for the transitions from $1^+_1$ and $1^+_2$ are in good 
agreements with the experimental data for the $1^+_1(T=0)$ at 12.1 MeV and
the $1^+_2(T=1)$ at 15.1 MeV. Therefore, we assigned 
the present $1^+_1$ and $1^+_2$ states to the observed $1^+_1(T=0)$
and $1^+_2(T=1)$ states, respectively.

\begin{table}
\caption{\label{tab:beta} 
Calculated log $ft$ values of the Gamov-Teller transitions 
for $\beta^+$ decays
compared with the experimental data for 
$^{12}$N($\beta^+$)$^{12}$C* 
taken from Ref.\protect\cite{isotopes}.}

\begin{center}
\begin{tabular}{cccccccc}
\multicolumn{3}{c}{Exp.} &  
\multicolumn{2}{c}{Theory(AMD)} \\
 Excitation energy (MeV) & $J^\pi_f$ & log$ft$ &
$J^\pi_f$ & log$ft$ \\
0  &  $0^+$ & 4.118$\pm$ 0.003 & 3.8 & $0^+_1$  \\
4.44  &  $2^+$ & 5.149$\pm$ 0.008 &  4.8 & $2^+_1$  \\
7.65  &  $0^+$ & 4.34$\pm$ 0.07 &   4.3  & $0^+_2$ \\
10.3  &  ($0^+$) & 4.36$\pm$ 0.18 & 4.7 &  $0^+_3$  \\
12.71  &  $1^+$ & 3.51$\pm$ 0.17 &  3.7  \\
  &  &  & 6.3  & $2^+_2$ \\
  &  &  & 5.0  & $2^+_3$ \\
\end{tabular}
\end{center}
\end{table}

\begin{table}
\caption{\label{tab:bm1} 
$M1$ transition strengths. 
The experimental data are 
taken from Ref.\protect\cite{isotopes}. The unit 
 is the nuclear unit, $(e\hbar/2Mc)^2$. The experimental 
excitation energies of the initial and final states are shown in 
parenthesis.
}

\begin{center}
\begin{tabular}{cccc}
\multicolumn{2}{c}{} & \multicolumn{2}{c}{$B(M1)$} 
 \\
$J^\pi_i$ & $J^\pi_f$ & Exp. & Theory(AMD) \\
$1^+_1$(12.1 MeV) & $0^+_1$(0 MeV) &   0.015 (0.002) & 0.011 \\
$1^+_1$(12.1 MeV) & $2^+_1$(4.4 MeV) & 0.0081 (0.0014) & 0.008 \\
$1^+_2$(15.1 MeV) & $0^+_1$(0 MeV) &  0.95 (0.02) & 0.17 \\
$1^+_2$(15.1 MeV) & $2^+_1$(4.4 MeV) & 0.068 (0.009) & 0.09 \\
$1^+_2$(15.1 MeV) & $0^+_2$(7.7 MeV) & 0.23 (0.04) & 0.014 \\
$1^+_2$(15.1 MeV) & $1^+_1$(12.1 MeV) & 3.6 (1.1) & 2.5 \\
\end{tabular}
\end{center}
\end{table}

\section{Discussions}\label{sec:discuss}

\subsection{Intrinsic structure}
\begin{figure}%----------------------------------------
\noindent
\epsfxsize=0.4\textwidth
\centerline{\epsffile{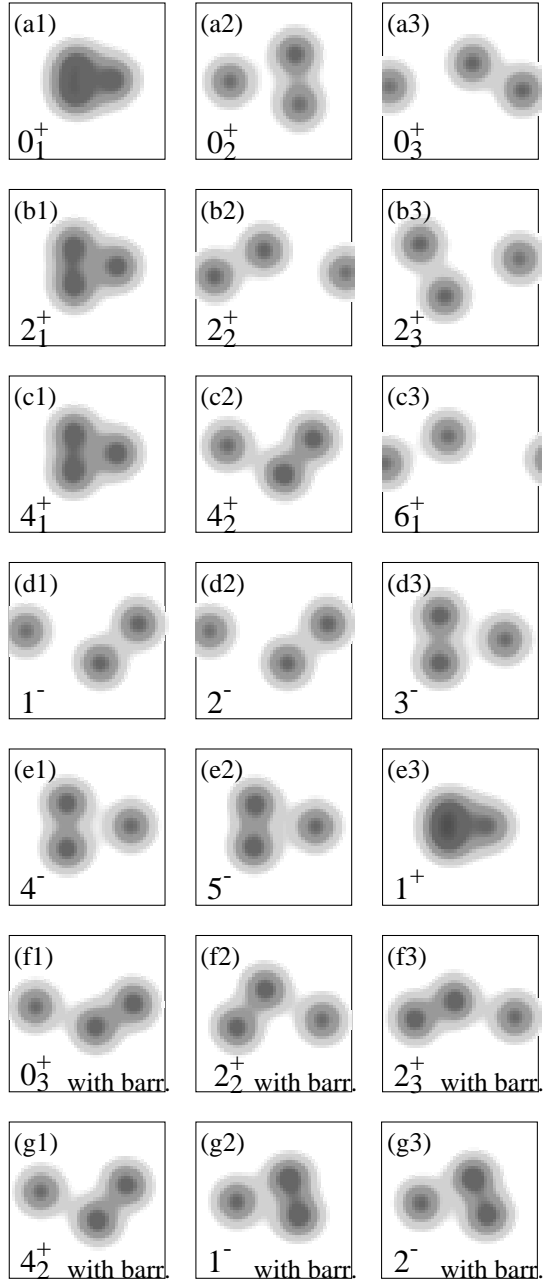}}
\caption{\label{fig:dense}
Mass density distributions of the intrinsic state 
$\Phi_{\rm AMD}({\bf Z}^{J\pi}_n)$ before the projection for the
states obtained by the VAP calculations with respect to the
$J^\pi_n$. Figures (f1)-(g3) are the mass density distributions 
of the states obtained with the artificial potential barrier.
The intrinsic system is projected 
onto a plane which contains the major axis of the 
intrinsic state.
The density is integrated along a transverse axis perpendicular to the
plane. The size of the frame box is 10 fm$\times$10 fm. 
}
\end{figure}

The present wave functions are given by the linear combination of 
the spin-parity eigen wave functions projected from 
23 AMD wave functions \{$\Phi_{\rm AMD}({\bf Z}_n^{J\pi})$\}. Here,
$\Phi_{\rm AMD}({\bf Z}_n^{J\pi})$ is the optimum AMD wave function,
which is obtained by the VAP calculation for the $J^\pi_n$ state.
Since each AMD wave function is written by a single Slater
determinant, it is easy to analyze its intrinsic structure.
Even though those bases are superposed by the diagonalization of 
the Hamiltonian
and norm matrices,  the $\Phi_{\rm AMD}({\bf Z}_n^{J\pi})$ is 
mostly the dominant component of the $J^\pi_n$ state after the 
diagonalization. 
Then we firstly discuss the intrinsic structure of the 
$\Phi_{\rm AMD}({\bf Z}_n^{J\pi})$.
In Figure \ref{fig:dense}, we show the density distributions of 
the intrinsic wave functions, $\Phi_{\rm AMD}({\bf Z}_n^{J\pi})$.
As seen in Fig.~\ref{fig:dense} (a1) and (e3), the $0^+_1$ and
the $1^+$ states have compact density distributions and no developed
cluster structure. 
In the $2^+_1$ and the $4^+_1$ states 
of the ground band built on the $0^+_1$ state, the 3$\alpha$ cluster
core appears(Figs.~\ref{fig:dense}-b1 and \ref{fig:dense}-c1). 
These states are considered to be 
the SU(3) limit 3$\alpha$ cluster states
because the spatial development of the
clustering is not remarkable. In the excited states, 
a variety of spatial configurations of the
developed 3$\alpha$ cluster structure appears. 
The $3\alpha$ in the $0^+_2$(Fig.~\ref{fig:dense}-a2) 
has an isosceles triangle configuration
which is close to an equilateral triangle.
The similar configurations of the $3\alpha$ are seen also in the 
$3^-_1$, $4^-_1$ and $5^-_1$ states
(Fig.~\ref{fig:dense}-d3,\ref{fig:dense}-e1,\ref{fig:dense}-e2). 
In the $0^+_3$ state(Fig.~\ref{fig:dense}-a3), the $3\alpha$ 
shows a rather linear-like configuration, where
the largest angle of vertices is larger than 120 degree. 
We found such the linear-like structure also in the $2^+_2$, $6^+_1$,
$1^-_1$ and $2^-_1$. In these states, a $\alpha$ cluster seems 
almost escaping. In Fig.~\ref{fig:dense}(f1)-(g3), 
we show the intrinsic structure obtained by the VAP 
with the artificial barrier potential as explained in \ref{sec:formulation}.
Although the $\alpha$ particles are more confined than 
the case without barrier, 
the developed 3$\alpha$ cluster structures are still formed in these
wave functions.

The spatial $3\alpha$ configurations of the $3^-$ and the $1^-$ states
in the present results are consistent with the energy  minimum states
in the isosceles configurations of the $3\alpha$ model 
space\cite{GCM}.
On the other hand, the present 
configuration of the $0^+_2$ state does not correspond to the
energy minimum but seems to be the second minimum state 
in the $3\alpha$ models space for the $0^+_2$
\cite{GCM}. 
After the superposition of the basis, the features of the $|0^+_2\rangle$ 
are similar to those of the $3\alpha$GCM calculations,
because various configurations of 
the $3\alpha$ cluster states mix in both the present results and 
$3\alpha$GCM calculations.

As explained in Sec.\ref{sec:formulation}, we 
describe the final wave functions $|J^\pi_n\rangle$
with a linear combination of the obtained wave functions, 
$P^{J\pi}_{MK}\Phi_{\rm AMD}({\bf Z}^{J_i\pi_i}_{k_i})$,
each of which is the spin-parity eigen wave functions projected from a 
AMD wave function.
The coefficients of Eq.~ \ref{eq:diago} 
are determined by the diagonalization of the 
Hamiltonian and norm matrices with respect to all the 
obtained AMD states.
In Table \ref{tab:overlap}, the amplitudes of the dominant components 
in the final wave functions $|J^\pi_n\rangle$ are listed.
In many cases, the dominant component in the $|J^\pi_n\rangle$ is the
$P^{J\pi}_{MK}\Phi_{\rm AMD}({\bf Z}^{J_i\pi_i}_{k_i})$ with 
$J_{i k_i}^{\pi_i}=J^\pi_n$, and
therefore, the $\Phi_{\rm AMD}({\bf Z}^{J\pi}_n)$ is regarded as the 
approximate intrinsic state of the $J^\pi_n$ state.

The $|0^+_1\rangle$, $|2^+_1\rangle$ and $|4^+_1\rangle$ 
in the ground band
contain about 90\% component of the dominant
$P^{J\pi}_{MK}\Phi_{\rm AMD}({\bf Z}^{J\pi}_k)$. 
In the $3\alpha$ cluster states, the amplitudes of the dominant 
component decrease mainly due to the mixing of various $3\alpha$
configurations. Especially, in the $|0^+_2\rangle$, the amplitude
of the $P^{0+}_{00}\Phi_{\rm AMD}({\bf Z}^{0+}_2)$ 
is only 0.49, and it contains other $3\alpha$ configurations
such as  $P^{0+}_{00}\Phi_{\rm AMD}({\bf Z}^{2+}_3)$.
The mixing of the different $3\alpha$ configurations
in the $|0^+_2\rangle$ 
enhances the loosely binding nature of the $3\alpha$ particles, and 
increases the $S$-wave component of the $^8$Be($0^+$)-$\alpha$ motion.
On the other hand, the dominant amplitude in the $|3^-_1\rangle$ is 0.82, 
which is large compared
with other $3\alpha$ cluster states. It may indicate that 
this state still has a nature of the rigid structure of the triangular 
$3\alpha$ configuration.   
The intrinsic state of the dominant component in the $|4^+_2\rangle$ and 
$|6^+_1\rangle$ is $\Phi_{\rm AMD}({\bf Z}^{6+}_1)$, where 
an $\alpha$ particle is almost escaping and locates at the distance 
$r\sim 7$ fm. Their stability against $\alpha$ decays should be carefully 
discussed to confirm the possible existence of these states.
Since the $|2^+_3\rangle$ is dominated by 
$P^{J\pi}_{MK}\Phi_{\rm AMD}({\bf Z}^{J_i\pi_i}_{k_i})$ with 
$J_{ik_i}^{\pi_i}=2^+_3$ and
$K=2$, this state is considered to be the band head state of 
a side-band $K^\pi=2^+$.
In the negative parity states, the intrinsic structures of the
$\Phi_{\rm AMD}({\bf Z}^{J\pi}_k)$'s with $J^\pi_k=3^-_1$, $4^-_1$, $5^-_1$ 
are similar to each other, and they construct the $K^\pi=3^-$ band; 
\{$3^-_1$, $4^-_1$, $5^-_2$\}.
On the other hand, the 
$\Phi_{\rm AMD}({\bf Z}^{J\pi}_k)$'s for $J^\pi_k=1^-_1$ and 
$J^\pi_k=2^-_1$ have
almost the same linear-like structures and form the $K^\pi=1^-$ band,
\{ $1^-_1$, $2^-_1$, $3^-_2$, $4^-_2$, $5^-_1$\}.
These negative parity bands, $K^\pi=3^-$ and $K^\pi=1^-$ were suggested also
in $3\alpha$GCM calculations.

\begin{table}
\caption{\label{tab:overlap} Overlap between 
$P^{J\pi}_{MK}\Phi_{\rm AMD}({\bf Z}^{J\pi}_k)$
and the final wave functions $|J^\pi_n\rangle$. The $|J^\pi_n\rangle$
is obtained by linear combination of 
$P^{J\pi}_{MK}\Phi_{\rm AMD}({\bf Z}^{J_i\pi_i}_{k_i})$ as explained in 
Eq.~ \protect\ref{eq:diago}.
$K$ is taken to be the same value as chosen in the VAP calculations. 
The overlap with the largest component in the $|J^\pi_n\rangle$
is also shown in case that the overlap with other components
is larger than 
$|\langle J^\pi_n| P^{J\pi}_{MK}
\Phi_{\rm AMD}({\bf Z}^{J\pi}_n)\rangle|^2$.
}
\begin{center}
\begin{tabular}{cccccccc}
{after}& {before}& & \\
diagonalization & diagonalization & overlap & overlap with\\ 
$J^\pi_n$ & $J^\pi_k$,$K$ & 
$|\langle J^\pi_n| P^{J\pi}_{MK}\Phi_{\rm AMD}({\bf Z}^{J\pi}_k)\rangle|^2$& 
the largest component \\
\hline
$\langle 0^+_1|$ & $0^+_1,0$ & 0.89 & \\
$\langle 0^+_2|$ & $0^+_2,0$ & 0.49 & 
$|\langle 0^+_2| P^{0+}_{00}\Phi_{\rm AMD}({\bf Z}^{2+}_{k=3})\rangle|^2=0.62$ \\
$\langle 0^+_3|$ & $0^+_3,0$ & 0.63 & \\
$\langle 1^+_1|$ & $1^+_1,0$ & 0.79 & \\
$\langle 2^+_1|$ & $2^+_1,0$ & 0.91 & \\
$\langle 2^+_2|$ & $2^+_2,0$ & 0.70 & \\
$\langle 2^+_3|$ & $2^+_3,0$ & 0.17  &  $|\langle 2^+_3| P^{2+}_{02}\Phi_{\rm AMD}({\bf Z}^{2+}_{k=3})\rangle|^2=0.51$ \\ 
$\langle 2^+_4|$ & $2^+_3,0$ & 0.27 &   $|\langle 2^+_4| P^{2+}_{00}\Phi_{\rm AMD}({\bf Z}^{2+}_{k=2});{\rm barrier}\rangle|^2=0.30$ \\ 
$\langle 4^+_1|$ & $4^+_1,0$ & 0.90 & \\
$\langle 4^+_2|$ & $4^+_2,0$ & 0.13 &   $|\langle 4^+_2| P^{4+}_{00}\Phi_{\rm AMD}({\bf Z}^{6+}_{k=1})\rangle|^2=0.47$ \\ 
$\langle 6^+_1|$ & $6^+_1,0$ & 0.57 &  $|\langle 6^+_1| P^{6+}_{00}\Phi_{\rm AMD}({\bf Z}^{2+}_{k=2})\rangle|^2=0.70$ \\ 
$\langle 1^-_1|$ & $1^-_1,1$ & 0.74 & \\
$\langle 2^-_1|$ & $2^-_1,1$ &  0.72 & $|\langle 2^-_1| P^{2-}_{0,-1}\Phi_{\rm AMD}({\bf Z}^{1-}_{k=1})\rangle|^2=0.79$ \\ 
$\langle 3^-_1|$ & $3^-_1,3$ & 0.82 & \\
$\langle 3^-_2|$ & $3^-_1,3$ & 0.10 & $|\langle 3^-_2| P^{3-}_{01}\Phi_{\rm AMD}({\bf Z}^{2-}_{k=1})\rangle|^2=0.69$ \\ 
$\langle 4^-_1|$ & $4^-_1,3$ & 0.72 & \\
$\langle 4^-_2|$ & $4^-_1,3$ & 0.26 & $|\langle 4^-_2| P^{4-}_{01}\Phi_{\rm AMD}({\bf Z}^{2-}_{k=1})\rangle|^2=0.59$ \\ 
$\langle 5^-_1|$ & $5^-_1,3$ & 0.18 & $|\langle 5^-_1| P^{5-}_{01}\Phi_{\rm AMD}({\bf Z}^{1-}_{k=1})\rangle|^2=0.77$ \\ 
$\langle 5^-_2|$ & $5^-_1,3$ &  0.76 & \\
\end{tabular}
\end{center}
\end{table}

In Table \ref{tab:tens1}, we show the expectation values of the 
total spin(${\bf J}$), the intrinsic spin(${\bf S}$), 
the angluar-momentum(${\bf L}$), 
and the principal quata of harmonic ocsillator(${N}^{\rm ho}$) 
for the spin-parity projected states
$P^{J\pi}_{MK}\Phi_{\rm AMD}({\bf Z}^{J\pi}_k)$. Here, the ocsillator quanta
$N^{\rm ho}$ is defined as, 
\begin{equation}
N^{\rm ho}\equiv \sum _i \left[\frac{{\bf p}^2_i}{4\hbar^2\nu}
+\nu{\bf r}^2_i -\frac{3}{2}\right ],
\end{equation}
where the width parameter $\nu$ is chosen to be the same value as that of the 
single-particle Gaussian wave packets.
We also show the expectation values of those operators for the
final wave functions $|J^\pi_n\rangle$ after the diagonalization 
in Table \ref{tab:tens2}.

Non-zero values of the intrinsic spins indicate the breaking of 
$3\alpha$ clustering. Due to the $3\alpha$ cluster structure, 
the $\langle {\bf S}^2\rangle$ values are 
almost zero in many states before the diagonalizaton
except for the $0^+_1$ and $1^+_1$ states 
as seen in Table\ref{tab:tens1}. 
The origin of the non-zero $\langle {\bf S}^2\rangle$ in the 
$0^+_1$ state is the mixing of the $p_{3/2}$ sub-shell closure component.
In case of the $1^+_1$ state, non-zero $\langle {\bf S}^2\rangle$ originates
in the intrinsic spin excitation. More detailed discussions of the 
$\alpha$-cluster breaking is given later.

The proton total spin $\langle {\bf J}^2_p\rangle$ 
and neutron total spin $\langle {\bf J}^2_n\rangle$ are the
smallest in the $0^+_1$ state. It is because of the significant 
$p_{3/2}$ sub-shell closure component. On the other hand, 
they are quite large in the $0^+_2$ and the $0^+_3$ states
compared with the $0^+_1$. This comes from the component of the 
orbital-angular momenta $L_p\ne 0$ and $L_n\ne 0$, 
which are caused by the
strong correlations of nucleons in the developed $\alpha$ clusters.
In case of the $0^+_2$ state, those values becomes larger in the 
$|0^+_2\rangle$ after the diagonalization of the basis than those in 
the $P^{0+}_{00}\Phi_{\rm AMD}({\bf Z}^{0+}_2)$
before the diagonalization.
It indicates that the further developed $3\alpha$ configurations
mix in the $|0^+_2\rangle$ through the superposition of the basis.

The oscillator quanta $N^{\rm ho}$
is equivalent to the principal quantum number of the 
spherical harmonic oscillator with the oscillation number 
$\omega=2\hbar\nu/ m$($m$ is the nucleon mass), and its expectation value 
indicates how large the mixing of higher shell components is in the expansion
of spherical harmonic oscillator basis. 
In the ground band members, $0^+_1$, $2^+_1$, $4^+_1$ states,
the values of the oscillator quanta for protons($\langle N^{\rm ho}_p\rangle$)
and neutrons($\langle N^{\rm ho}_n\rangle$) nearly equal to the minimum value
4, which correspond to the $0\hbar\omega$ states. 
On the other hand, those for the spatially developed 
$3\alpha$ cluster states such as the $0^+_2$, $0^+_3$, $1^-_1$ and $3^-_1$
are much larger than 4. These cluster states
should be written by linear combinations of a large number of the 
higher shell configurations in the expansion of the harmonic oscillator
basis. Therefore, it is natural that 
these cluster states can not be 
described even by the large-basis shell model calculations
\cite{navratil03}.

\begin{table}
\caption{\label{tab:tens1} 
Expectation values of the squared total spin(${\bf J}^2$), the squared 
intrinsic spin(${\bf S}^2$), 
the squared angular-momentum(${\bf L}^2$), 
and the principal quanta of harmonic oscillator(${N}^{\rm ho}$) 
for the spin-parity projected states
$P^{J\pi}_{MK}\Phi_{\rm AMD}({\bf Z}^{J\pi}_k)$ before the diagonalization. 
$K$ is chosen to be the same
value as used in the VAP calculations. The calculated values for 
protons(${\bf J}_p$, ${\bf S}_p$, ${\bf L}_p$, ${N}^{\rm ho}_p$), 
neutrons(${\bf J}_n$, ${\bf S}_n$, ${\bf L}_n$, ${N}^{\rm ho}_n$),
and the total system(${\bf S}$, ${\bf L}$)
are shown.
}
\begin{center}
\begin{tabular}{cccccccccccc}
$J^\pi_k$ &  $|K|$ &  $\langle{\bf J}^2_p\rangle$ & $\langle{\bf J}^2_n\rangle$
& $\langle{\bf S}^2\rangle$ & $\langle{\bf S}^2_p\rangle$ 
& $\langle{\bf S}^2_n\rangle$ 
& $\langle{\bf L}^2\rangle$ & $\langle{\bf L}^2_p\rangle$ 
& $\langle{\bf L}^2_n\rangle$
& $\langle{N}^{\rm ho}_p\rangle$ & $\langle{N}^{\rm ho}_n\rangle$ \\
\hline
$0^+_1$ & 0 & 	0.8 	&	0.8 	&	1.6 	&	0.8 	&	0.8 	&	1.6 	&	1.5 	&	1.5 	&	4.1 	&	4.1 \\
$0^+_2$ & 0 & 	6.5 	&	6.5 	&	0.1 	&	0.0 	&	0.0 	&	0.1 	&	6.5 	&	6.5 	&	6.6 	&	6.5 \\
$0^+_3$ & 0 &  15.1 	&	15.1 	&	0.0 	&	0.0 	&	0.0 	&	0.0 	&	15.1 	&	15.1 	&	14.3 	&	14.3 \\
$1^+_1$ & 0 &  2.3 	&	0.9 	&	3.4 	&	2.0 	&	1.0 	&	3.6 	&	2.3 	&	1.6 	&	4.3 	&	4.1 \\
$2^+_1$ & 0 & 4.3 	&	4.2 	&	0.6 	&	0.3 	&	0.3 	&	6.2	&	4.3 	&	4.3 	&	4.4 	&	4.4 \\
$2^+_2$ & 0 & 16.2 	&	16.2 	&	0.0 	&	0.0 	&	0.0 	&	6.0 	&	16.2 	&	16.2 	&	14.1 	&	14.0 \\
$2^+_3$ & 0 & 12.3 	&	12.3 	&	0.0 	&	0.0 	&	0.0 	&	6.2 	&	12.3 	&	12.3 	&	10.4 	&	10.3 \\
$4^+_1$ & 0 & 7.2 	&	7.1 	&	0.3 	&	0.1 	&	0.1 	&	19.3 	&	7.0 	&	6.9 	&	4.5 	&	4.5 \\
$4^+_2$ & 0 & 12.8 	&	12.8 	&	0.1 	&	0.0 	&	0.0 	&	20.0 	&	12.8 	&	12.8 	&	7.6 	&	7.6 \\
$1^-_1$ & 1 & 11.9 	&	11.9 	&	0.0 	&	0.0 	&	0.0 	&	2.0 	&	11.9 	&	11.9 	&	11.3 	&	11.2 \\
$2^-_1$ & 1 & 13.9 	&	13.9 	&	0.0 	&	0.0 	&	0.0 	&	6.0 	&	13.9 	&	13.9 	&	12.4 	&	12.3 \\
$3^-_1$ & 3 & 9.7 	&	9.7 	&	0.1 	&	0.0 	&	0.1 	&	12.0 	&	9.7 	&	9.6 	&	6.4 	&	6.3 \\
$4^-_1$ & 3 &	11.3 	&	12.0 	&	0.1 	&	0.0 	&	0.0 	&	19.9 	&	11.3 	&	11.1 	&	6.3 	&	6.3 \\
\end{tabular}
\end{center}
\end{table}

\begin{table}
\caption{\label{tab:tens2} 
Expectation values of the squared total spin(${\bf J}^2$), the squared 
intrinsic spin(${\bf S}^2$), 
the squared angular-momentum(${\bf L}^2$), 
and the principal quanta of harmonic oscillator(${N}^{\rm ho}$) 
for the final wave functions $|J^\pi_n\rangle$
obtained by the superposition of the wave functions
after the diagonalization.
The calculated values for 
protons(${\bf J}_p$, ${\bf S}_p$, ${\bf L}_p$, ${N}^{\rm ho}_p$), 
neutrons(${\bf J}_n$, ${\bf S}_n$, ${\bf L}_n$, ${N}^{\rm ho}_n$),
and the total system(${\bf S}$, ${\bf L}$)
are shown.
}
\begin{center}
\begin{tabular}{ccccccccccc}
$|J^\pi_k\rangle $ &   $\langle{\bf J}^2_p\rangle$ & $\langle{\bf J}^2_n\rangle$
& $\langle{\bf S}^2\rangle$ & $\langle{\bf S}^2_p\rangle$ 
& $\langle{\bf S}^2_n\rangle$ 
& $\langle{\bf L}^2\rangle$ & $\langle{\bf L}^2_p\rangle$ 
& $\langle{\bf L}^2_n\rangle$
& $\langle{N}^{\rm ho}_p\rangle$ & $\langle{N}^{\rm ho}_n\rangle$ \\
\hline
$0^+_1$ &	1.7 	&	1.7 	&	1.2 	&	0.6 	&	0.6 	&	1.2 	&	2.2 	&	2.2 	&	4.4 	&	4.4 \\
$0^+_2$ &	8.0 	&	8.0 	&	0.6 	&	0.3 	&	0.3 	&	0.6 	&	8.3 	&	8.3 	&	8.4 	&	8.3 \\
$0^+_3$ &	14.4 	&	14.4 	&	0.2 	&	0.1 	&	0.1 	&	0.2 	&	14.5 	&	14.5 	&	14.0 	&	13.9 \\
$1^+_1$ &	1.8 	&	1.5 	&	3.9 	&	1.6 	&	1.4 	&	4.0 	&	2.1 	&	1.9 	&	4.2 	&	4.2 \\
$1^+_2$ &	1.5 	&	1.9 	&	2.2 	&	1.3 	&	1.5 	&	2.3 	&	1.8 	&	2.0 	&	4.2 	&	4.2 \\
$2^+_1$ &	4.7 	&	5.0 	&	0.5 	&	0.3 	&	0.2 	&	6.1 	&	4.7 	&	5.0 	&	4.8 	&	4.8 \\
$2^+_2$ &	16.1 	&	16.1 	&	0.0 	&	0.0 	&	0.0 	&	6.0 	&	16.1 	&	16.1 	&	13.9 	&	13.9 \\
$2^+_3$ &	11.3 	&	11.8 	&	0.3 	&	0.2 	&	0.1 	&	6.0 	&	11.3 	&	11.7 	&	10.1 	&	10.1 \\
$2^+_4$ &	15.0 	&	15.0 	&	0.1 	&	0.0 	&	0.0 	&	6.0 	&	15.0 	&	15.0 	&	13.0 	&	12.9 \\
$4^+_1$ &	8.0 	&	8.1 	&	0.2 	&	0.1 	&	0.1 	&	19.5 	&	7.9 	&	8.0 	&	5.1 	&	5.1 \\
$4^+_2$ &	21.2 	&	21.3 	&	0.0 	&	0.0 	&	0.0 	&	20.0 	&	21.2 	&	21.3 	&	15.5 	&	15.4 \\
$1^-_1$ &	9.8 	&	9.9 	&	0.1 	&	0.0 	&	0.0 	&	2.1 	&	9.8 	&	9.9 	&	9.4 	&	9.3 \\
$2^-_1$ &	11.4 	&	11.4 	&	0.1 	&	0.0 	&	0.0 	&	6.0 	&	11.4 	&	11.4 	&	9.9 	&	9.8 \\
$3^-_1$ &	10.8 	&	10.6 	&	0.1 	&	0.1 	&	0.1 	&	12.0 	&	10.8 	&	10.5 	&	7.4 	&	7.3 \\
$3^-_2$ &	14.1 	&	14.1 	&	0.1 	&	0.1 	&	0.1 	&	12.0 	&	14.0 	&	14.1 	&	10.9 	&	10.8 \\
$4^-_1$ &	13.0 	&	12.9 	&	0.1 	&	0.0 	&	0.1 	&	19.9 	&	13.0 	&	12.9 	&	7.9 	&	7.8 \\
$4^-_2$ &	15.3 	&	15.4 	&	0.1 	&	0.0 	&	0.0 	&	20.0 	&	15.3 	&	15.4 	&	10.4 	&	10.3 \\
\end{tabular}
\end{center}
\end{table}

\subsection{Breaking of $\alpha$ clusters}

In the present framework, the existence of $\alpha$ clusters is not
assumed, but the formation and breaking of the clusters are automatically 
incorporated in the energy variation if such the structures are favored.
We can estimate the $\alpha$ breaking components with the non-zero 
values of the squared intrinsic spins, $\langle {\bf S}^2 \rangle$,
listed in Table \ref{tab:tens1} and \ref{tab:tens2}.

Let us first discuss the cluster and non-cluster components 
in the wave functions by analyzing the results shown in Table 
\ref{tab:tens1} before the diagonalization.
The states 
except for the $0^+_1$, $2^+_1$, $4^+_1$ and $1^+_1$ have
almost zero values of $\langle {\bf S}^2 \rangle$
because of the well-developed $3\alpha$ cluster structures.
It means that the $\alpha$ clusters are not broken in these cluster states.
On the other hand, 
$\langle {\bf S}^2 \rangle=1.6$ in the $0^+_1$ state indicates the
significant $\alpha$ breaking component. It is natural because  
$p_{3/2}$ sub-shell closed configuration is favored in the ground state. 
If the ground state is the
pure $p_{3/2}$ sub-shell closed state, the expectation values of the 
squared total spins for protons($\langle {\bf J}^2_p \rangle$) 
and neutrons($\langle {\bf J}^2_n \rangle$) should be zero. Considering 
the non-zero values($\langle {\bf J}^2_p \rangle=\langle 
{\bf J}^2_n \rangle=0.8$), the $0^+_1$ state is considered to be 
a mixture of the $SU(3)$-limit 3$\alpha$ state and the 
$p_{3/2}$-shell closed state.
In other words, the
$3\alpha$ clustering is partially broken due to the spin-orbit force in the
ground state. 
The $1^+$ is the non 3$\alpha$ state with intrinsic spin excitations 
as seen in its spin magnitude($\langle {\bf S}^2 \rangle=3.4$). 
It is consistent with the
fact that the $J^\pi=1^+$ 
is unnatural in a $3\alpha$ system.

Next we look into the $\alpha$ breaking in the $|J^\pi_n\rangle$
after the diagonalization. By comparing the values in Table \ref{tab:tens2}
with those in Table \ref{tab:tens1}, we found that 
the $\alpha$ breaking component in the original wave function of the
ground state mixes into the $|0^+_2\rangle$ and $|0^+_3\rangle$ because
the $\langle {\bf S}^2\rangle$ values of the
$|0^+_2\rangle$ and $|0^+_3\rangle$ become large due to the diagonalization.
In the $0^+_2$ state,
the angular momentum 
$\langle {\bf L}_{p(n)}^2 \rangle$ for protons(neutrons) increases as well as
the intrinsic spins $\langle {\bf S}^2 \rangle$ due to the
superposition of the wave functions. 
It indicates
that not only the $\alpha$-breaking component but also the 
component of further developed $3\alpha$ clustering increases
in the $0^+_2$ state through the diagonalization. 

As mentioned before, the GT transition strengths are
good probes to investigate the mixing of cluster and non-cluster components,
because the transitions to the ideal $3\alpha$ cluster states
are forbidden, and therefore, the strengths directly reflect the
$\alpha$ breaking component. In the experimental observations, 
$\beta$ decays from $^{12}$N to the ground state of $^{12}$C
are rather strong as shown in Table \ref{tab:beta}. 
Even for the transitions to the excited states, $0^+_2$ and $0^+_3$,
the significant strengths of the $\beta$ decays were observed.
These facts indicate that the $\alpha$ breaking component is contained 
in the $0^+_2$ and $0^+_3$ states as well as the $0^+_1$ state.
In the present study, we found that 
the $\alpha$ breaking is significantly included in the original $0^+_1$ 
wave function due to the spin-orbit force, while the original 
$0^+_2$ and $0^+_3$ wave functions before the diagonalization 
have the well-developed $3\alpha$ structure
but no $\alpha$ breaking.
When we calculated the GT transitions for the original wave functions
before the superposition of the basis, we obtained 
almost forbidden GT transitions into the $0^+_2$ and $0^+_3$
states because of the $3\alpha$ cluster structures,
and failed to reproduce the experimental log$ft$ values.
The $\alpha$ breaking components are 
slightly contained in the $|0^+_2\rangle$ and $|0^+_3\rangle$ after the
diagonalization through the mixing of the original $0^+_1$ wave function,
which includes the $p_{3/2}$ sub-shell closed configuration.
As a result of the mixing of the 
$\alpha$ breaking components, the log$ft$ values 
are well reproduced by the present wave functions after the diagonalization. 

The $\alpha$ breaking gives important effects on various properties
of the ground and excited states of $^{12}$C as well as the GT transition
strengths.
As mentioned before, the level spacing between 
the $0^+_1$ and $2^+_1$ states is as large as the experimental data because 
the ground state gains the spin-orbit force due to the $\alpha$
breaking. This large level spacing can not be described by the 
$3\alpha$ cluster models.
The $\alpha$ breaking component of the ground state should affect
also the properties of the excited $0^+$ states. 
Within the $3\alpha$ model space, the ground state is almost the $SU(3)$-limit
$3\alpha$ state, while the excited $0^+_2$ state is an optimum solution
in the model space orthogonal to the $3\alpha$ ground state.
It means that the compact $3\alpha$ state is forbidden for the $0^+_2$ state
to satisfy the orthogonality to the ground state.
In the present work, since the ground state contains the $\alpha$ 
breaking component, the $SU(3)$-limit $3\alpha$ state is partially allowed 
for the $0^+_2$ state. We consider that this is one of the reasons for the
smaller radius of the $0^+_2$ state in the present calculations
than the values obtained by the $3\alpha$ calculations. 
The reduction of the $0^+_2$  size
due to the non-$3\alpha$ component in the ground state 
should be important also in the description of the 
experimental value of the $E2$ strength.
The experimental data of $B(E2;0^+_2\rightarrow 2^+_1$) is usually
underestimated by  the $3\alpha$ calculations.
In the present study, we can describe the increase of 
$B(E2;0^+_2\rightarrow 2^+_1$) by the effect of the size reduction, which 
enlarges the overlap of the $0^+_2$ with the $2^+_1$.  
Since the present $B(E2;0^+_2\rightarrow 2^+_1$) value is 
larger than the experimental one, the effect of the $\alpha$ breaking 
might be overestimated.

Let us discuss the $\alpha$ breaking effect on the
inelastic form factor to the $0^+_2$ state.
Funaki {\em et al.} showed that 
the maximum value of the inelastic form factor 
is sensitive to the spatial extension of the $0^+_2$ state.
In their analysis based on the $3\alpha$ condensate wave function, 
the magnitude of the form factor decreases with the enhancement of 
the $0^+_2$ size. They 
succeeded to reproduce the experimental data of the inelastic 
form factor by the $3\alpha$ condensate wave function 
with a large radius $R_m=3.8$ fm.
On the other hand, as shown in Fig.~\ref{fig:form}, 
the present results well 
reproduce the maximum value of the form factor, though the root-mean-square
radius of the $0^+_2$ is $R_m=3.3$ fm which is smaller than that of 
their calculations. 
The $E0$ strengths should be sensitive also to the
structure of the ground state.
If the ground state contains the $\alpha$ breaking component, 
the inelastic transition strength to the excited $3\alpha$ state 
should decrease, because the non-$3\alpha$ component reduces the
transition overlap between the ground state and 
the $\alpha$ cluster states in general.
This is one of the reasons for the suppression of the 
inelastic transitions in the present calculations, which 
reproduce the magnitude of the inelastic form factor for 
the $0^+_1\rightarrow 0^+_2$. 

It should be stressed that the $\alpha$ breaking component 
in the ground state is directly reflected 
in the inelastic transition strengths from the ground state, 
and also can have an influence 
on the structure of the excited $0^+$ states.

\subsection{$3\alpha$ cluster features}

In many theoretical works, 
the $3\alpha$ cluster features of $^{12}$C 
have been discussed by many groups for a long time. 
Especially, the $0^+$ and the $2^+$ states above the $3\alpha$ threshold
energy have attracted great interests. Recently, 
there are some experimental reports on the $0^+$ and $2^+$ around $E_x=10$
MeV\cite{john03,fynbo03,itoh04,diget05}. 
However, the level structure and the assignment of these resonances
have not been clarified yet. In the present work, various $3\alpha$ 
cluster structures 
appear in the excited states of $^{12}$C. 
We, here, analyze the structure of the $0^+$, the $2^+$ and the $4^+$ states
by extracting the $^8$Be($0^+$)+$\alpha$ components 
and estimate the $\alpha$ decay widths with the method of
reduced width amplitudes. We give a discussion of the 
level assignment and the band structure later in \ref{sec:level}. 

We extracted the reduced width amplitudes for the $^8$Be($0^+$)+$\alpha$, 
and calculated the spectroscopic factors($S$) 
and cluster probabilities. 
For simplicity, we assumed the $(0s)^4$ wave function for the 
$\alpha$ particle and $SU(3)$-limit cluster wave function 
for the $^8$Be($0^+$) by using the same width parameter $\nu$ as 
the present AMD wave function of the $^{12}$C.
The definitions and the detailed method of the calculations 
are explained in Ref.\cite{ENYO-be12}.

\begin{figure}%----------------------------------------
\noindent
\epsfxsize=0.49\textwidth
\centerline{\epsffile{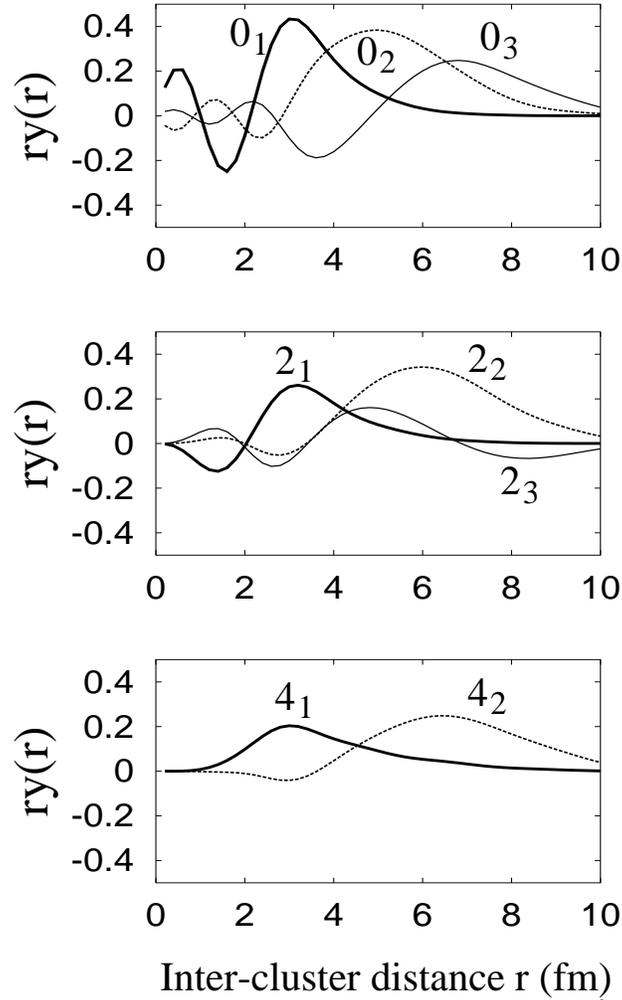}}
\caption{\label{fig:yl}
$\alpha$-reduced width amplitudes $ry(r)$ of the $0^+$, $2^+$ and $4^+$ states
for the [$^8$Be$(0^+)\otimes l]_{J}(l=J)$ channel. 
}
\end{figure}

Figure \ref{fig:yl} shows the reduced width amplitudes $ry_L(r)$ of the 
$0^+$, $2^+$ and $4^+$ states. The calculated $S$ factors and cluster 
probabilities are listed in Table \ref{tab:sfac}.
In the $0^+_1$, $2^+_1$ and $4^+_1$ states, 
the node numbers of the cluster wave functions are 2, 1 and 0, respectively, 
which correspond to the lowest allowed node numbers for 
the $^8$Be-$\alpha$ relative motion.
In the higher $J^\pi$ states, the node number increases one by one 
and the position of the largest peak shifts toward
the large distance region. 
The $S$ factors in Table \ref{tab:sfac} indicate the 
spatial developments of the $^8$Be($0^+$)+$\alpha$ clustering.
The $S$ factor is the largest in
the $0^+_2$ state. Among three $2^+$ states, $S$ is the largest
in the $2^+_2$ state. 
In the $0^+_3$ state,
the $S$ factor is not so large even though this state is the developed
$3\alpha$ cluster state.   
It is because the main component of the $0^+_3$ is a rather
linear-like $3\alpha$, which results in the reduction of the 
$^8$Be($0^+$)+$\alpha$ component because of 
the large mixing of the $^8$Be($2^+$)+$\alpha$ with 
$D$-wave relative motion. 
These behavior of the cluster wave functions $ry_L(r)$ for the
$0^+_1$, $0^+_2$, $0^+_3$, $2^+_1$, $2^+_2$ and $2^+_3$ corresponds
well to the $3\alpha$GCM calculations by Uegaki {\em et al.}\cite{GCM}, 
though the
present amplitudes are slightly smaller than those of the $3\alpha$GCM
results. 
Therefore, the basic features of the $3\alpha$ cluster components
in the present results are considered to be similar 
to those of the $3\alpha$GCM results. 

The calculated reduced widths($\theta_\alpha^2(a)$) are shown 
in Table \ref{tab:sfac}. The large values of $\theta_\alpha^2(a)$ in
the $0^+_2$, $0^+_3$ and $2^+_2$ states 
imply the spatial development of the $\alpha$ cluster
in these states. We estimated the  
$^8$Be($0^+$)+$\alpha$ decay width with the method of 
reduced width amplitudes in the same way as in Ref.\cite{ENYO-be12}. 
In the present estimation, we temporary assigned the $0^+_3$ 
and the $2^+_2$ to the observed states around $E_x=10\sim 11$MeV 
and chose the energy of the relative 
motion $E_\alpha$ to be 3 MeV. 
The calculated $\alpha$ decay widths for the $0^+_3$ and the $2^+_2$
are 400 keV and 460 keV, respectively. In these energy region, the
partial width for the $^8$Be($0^+$)+$\alpha$ decay is expected to dominate
the total $\alpha$ decay widths as shown in Ref.\cite{GCM}.

\begin{table}
\caption{\label{tab:sfac} 
The calculated reducued $\alpha$-decay widths($\theta^2$),
the amplitudes ($ry_L(r)$), 
$S$ factors, cluster probabilities, $\alpha$-decay widths
for the $^8$Be($0^+$)+$\alpha$ channel.
The assumed 
excitation energies $E_x$ for the
$0^+_3$ and the $2^+_2$ states are shown in parenthesis.
The calculated $\alpha$ decay widths are obtained with the experimental energy
$E_\alpha$ of the relative motion between $^8$Be($0^+$) and $\alpha$.
The experimental widths\protect\cite{isotopes} are also shown. 
}

\begin{center}
\begin{tabular}{cccccccccc}
$J^\pi$ & $E_x$(MeV) & $E_\alpha$ & channel radius &  & reduced width & $S$-factor & Cluster probability & $\Gamma_{\rm cal}$(keV) & $\Gamma_{\rm exp}$(keV) \\ 
      &            &             & $a$(fm) & $|ay(a)|^2$(fm$^2$)
& $\theta_\alpha^2(a)$ & & & &  \\
$0^+_1$	&	0	&		&	4.0	&	6.3E-02	&	0.08 	&	0.31 	&	0.43 	&		&		\\
$0^+_2$	&	7.654	&	0.38	&	6.0	&	9.6E-02	&	0.19 	&	0.40 	&	0.43 	&	0.04	& 8.5e-3		\\
$0^+_3$	&	(10.3)	&	(3)	&	7.0	&	6.0E-02	&	0.14 	&	0.19 	&	0.19 	&	400	& (3000)		\\
$2^+_1$	&	4.44	&		&	4.0	&	3.3E-02	&	0.04 	&	0.12 	&	0.47 	&		&		\\
$2^+_2$	&	(10.3)	&	(3)	&	6.0	&	1.2E-01	&	0.23 	&	0.33 	&	0.35 	&	460	& 		\\
$2^+_3$	&		&		&	8.0	&	4.0E-03	&	0.01 	&	0.06 	&	0.11 	&		&		\\
$4^+_1$	&	14.08	&	6.81	&	4.0	&	2.1E-02	&	0.03 	&	0.08 	&	0.52 	&	40	&	258	\\
$4^+_2$	&		&		&	7.0	&	5.0E-02	&	0.12 	&	0.18 	&	0.19 	&		&		\\
\end{tabular}
\end{center}
\end{table}

\subsection{Assignments of $0^+$ and $2^+$ resonances}\label{sec:level}

The $0^+$ and $2^+$ states around $E_x=10$ MeV have been 
experimentally and theoretically studied for a long time.
We here consider the correspondence of the
present results with other theoretical calculations and 
the experimental reports.

There exists the broad resonance at $E_x=$10.3 MeV with the width 
$\Gamma\sim 3$ MeV, which has been ambiguously assigned to a $0^+$ state
\cite{isotopes}. Because of the existence of this broad resonance,
there remain uncertainties on the spectroscopy in this energy region.
In the inelastic $^6$Li and $\alpha$ scattering\cite{john03,eyrich87}, 
a broad $0^+$ state has 
been observed around $E_x=10$ MeV. Its energy and width are consistent with the
broad $0^+$ at $E_x=$10.3 MeV.
The $0^+$ state around $10\sim 11$ MeV has been also observed in the recent 
experiments of the $\beta$-delayed $3\alpha$ particle 
decay from $^{12}$B and $^{12}$N\cite{fynbo03,diget05}. 
In the argument in \cite{diget05}, interference between this state and the
$0^+_2$(7.65 MeV) has so significant influence on the spectrum and it
makes difficult to determine the resonance energy.
Recently, the $2^+_2$ state has been discovered at $E_x\sim 10$ MeV 
by the inelastic $\alpha$ scattering\cite{itoh04}. 
The spectra of this state are overlapped by the broad peak of 
the $0^+$ at 10.3 MeV, and the fitting parameter for the
$2^+_2$ is $E_x=9.9\pm 0.3$ MeV and 
$\Gamma=1.0\pm 0.3$ MeV.

The $0^+$ state above the $3\alpha$ threshold has been long discussed 
also in many theoretical studies. Morinaga {\em et al.}\cite{morinaga66}
proposed a linear-chain $3\alpha$ structure(LCS) 
of the $0^+_2$ at 7.65 MeV.
The possibility of the LCS has been discussed based 
on the analysis of the $\alpha$ decay widths 
by Suzuki {\em et al.}\cite{suzuki72}, and it was argued that 
the LCS is inappropriate to explain the large experimental widths of  
the $0^+$(7.65 MeV) and the $0^+$(10.3 MeV).
In the microscopic and semi-microscopic $3\alpha$ cluster model 
calculations\cite{RGM,GCM,kurokawa04,funaki03}, 
it was found that the $0^+_2$ has the developed $3\alpha$ structure with
the dominant [$^8$Be$(0^+)\otimes l=0]_{J=0}$ and has 
the nature of the weakly interacting $3\alpha$ particles.
This is quantitatively consistent with the features of $3\alpha$ clustering 
in the $0^+_2$ state of the present results.
Recently, the new interpretation of the $0^+_2$ as 
a $3\alpha$-cluster gas was proposed \cite{Tohsaki01,funaki03}. 
Also in the present study, we found the  
dilute $3\alpha$ gas aspects of 
the $0^+_2$, though its radius is 
somewhat smaller than that of the $3\alpha$ calculations. 
In these theoretical studies, it is considered that 
the $3\alpha$ structure of the $0^+_2$ has been qualitatively
established.

The $0^+_3$ state 4 MeV above the $0^+_2$ state 
was predicted in the $3\alpha$GCM\cite{GCM}, 
$3\alpha$RGM+CSM\cite{Pichler97}, 
and $3\alpha$OCM+CSM\cite{kurokawa04} calculations. In the comparison 
with the experimental data of the $0^+$(10.3 MeV,$\Gamma_{\rm exp}$=3 MeV), 
the theoretical energy is rather higher than the observed data, 
and the calculated width $\Gamma_{\rm cal}\sim 1$ MeV is narrower than 
the $\Gamma_{\rm exp}=3$ MeV. 
Also in the present results, we obtained the $0^+_3$ state 
a few MeV above the $0^+_2$ state.
The structure of this state is similar to the $0^+_3$ state of 
the $3\alpha$GCM calculations. 
As mentioned before, the dominant $3\alpha$ configuration of the $0^+_3$ is 
rather linear-like as well as the $3\alpha$GCM calculations, where
the $0^+_3$ state is dominated by [$^8$Be$(2^+)\otimes l=2]_{J=0}$ and 
has less [$^8$Be$(0^+)\otimes l=0]_{J=0}$ component. 
This results in the suppression 
of the $\alpha$ decay width. Comparing with the 
observed broad width, it seems that the theoretical $0^+_3$ state
should not be directly assigned to the experimental 
$0^+$(10.3 MeV,$\Gamma_{\rm exp}$=3 MeV).
Recently, Kurokawa {\em et al.} predicted a new broad $0^+$ 
resonance below the predicted $0^+_3$ state. 
It may suggest that another broad resonance might coexist with the
predicted narrower $0^+_3$ state in this energy region. 
The possible coexistence of two $0^+$ resonances in the 
broad $0^+$ spectra at 10.3 MeV with $\Gamma_{\rm exp}=3$ MeV
has not been experimentally excluded.
By speculating the coexistence of the 
two $0^+$ states in this energy region, 
we conjecture that 
the $0^+_3$ in the present result may correspond to the narrower
state of the two $0^+$ resonances. 
There exists some experimental information concerning the
total strengths of the transitions to the broad $0^+$(10.3 MeV).
The inelastic 
$^6$Li and $\alpha$ scattering data imply that the isoscalar monopole 
transition strengths to the $0^+$(10.3 MeV) is 
30-100\% of that to the $0^+$(7.65 MeV), while in the present results
it is 10\% of the strength to the $0^+_2$ state.
On the other hand, the calculated $B(GT)$ for the GT transition from 
$^{12}$N to the $0^+_3$ is half of the experimental strengths for the
transition to the $0^+$(10.3 MeV) (see Table \ref{tab:beta}).
In order to clarify the properties of the $0^+_3$ state, 
more precise experimental data and further theoretical studies are required.

Next, we discuss the $2^+_2$ state. The cluster models have predicted the 
$2^+_2$ state around $E_x=10$ MeV as a member of the rotational band built
on the $0^+_2$ state\cite{RGM,GCM,kurokawa04,funaki05}. 
The recent experimental report on the $2^+_2$ at $9.9\pm 0.3$ MeV
with $\Gamma=1.0\pm 0.3$\cite{itoh04} is consistent with 
the theoretical energy and width of the predicted $2^+_2$ 
in Refs.\cite{kurokawa04,funaki05} and also with the present results. 
The theoretical $2^+_2$ state is dominated by 
[$^8$Be$(0^+)\otimes l=2]_{J=2}$ as shown in Table \ref{tab:sfac}
as well as in the $3\alpha$ models\cite{GCM,kurokawa04,funaki05}.
Therefore, in the weak coupling picture of a $^8$Be+$\alpha$ system,
the $2^+_2$ state is regarded as the rotational member upon the 
$0^+_2$ state, which has the dominant [$^8$Be$(0^+)\otimes l=0]_{J=0}$.
The $E2$ strength from the $2^+_2$ to the $0^+_2$ is large as
$B(E2;2^+_2\rightarrow 0^+_2)=100$ e$^2$fm$^4$ in the present calculations.
However, by analyzing the $E2$ transitions and the intrinsic structure
of the $0^+$ and the $2^+$ states,  
we can extract an alternative interpretation for the band
assignment of the $2^+_2$.
Namely, the $2^+_2$ and the $0^+_3$ can be the rotational band members of the 
linear-like $3\alpha$ structure. 
In fact, we obtained the remarkably large $B(E2)$ values,
$B(E2;2^+_2\rightarrow 0^+_3)=310$ e$^2$fm$^4$. Moreover, 
the dominant intrinsic wave functions of the $2^+_2$ and the $0^+_3$ 
are quite similar to each other. They show 
the linear-like $3\alpha$ structure, 
where the largest angle of vertices of the triangle $3\alpha$ configuration
is larger than 120 degree (Figs.~\ref{fig:dens}-a3 and \ref{fig:dens}-b2). 
The overlaps of the $|2^+_2\rangle$ and $|0^+_3\rangle$ 
with the wave function $P^\pm_{MK}\Phi_{\rm AMD}({\bf Z}^{0+}_3)$
projected from the single AMD wave function
(Fig.~\ref{fig:dens}-a3) are more than 60\%.
As a result of the similar intrinsic structure with 
the large deformation, $B(E2;2^+_3\rightarrow 0^+_3)$ is extremely large, 
and the shape of the density $\rho(r)$ of these two states 
are quite similar to each other(Fig.~\ref{fig:dens}). 
Consequently, in a collective 
picture, we consider that the $0^+_3$ and the $2^+_2$ states 
are regarded as the members of 
the rotational band, which is formed by the deformed intrinsic state
with the linear-like $3\alpha$ structure.
We expect that the real $2^+_2$ state may have an intermediate rotational 
feature of this strong coupling aspect built on the $0^+_3$ state and 
the weak coupling one built on the $0^+_2$ state.
Unfortunately, the experimental information for the $2^+_2$ and the 
$0^+_3$ state is too poor to justify the band assignment of the $2^+_2$ state.

\section{Summary}\label{sec:summary}
We investigated structure of the ground and excited states of 
$^{12}$C. The present theoretical method is based on the VAP calculations
in the framework of AMD, which can describe both 
cluster and shell-model-like aspects. The present results systematically 
reproduce the various observed data such as the energy levels, electric 
and $\beta$ transition strengths.
It was found that the ground state is the mixture of 
the shell-model-like state and the $3\alpha$ cluster state, 
while the developed $3\alpha$ cluster structures appear in the
excited states. We also obtained the $1^+$ states with 
non-$3\alpha$ cluster structure.

We discussed the $\alpha$ breaking components. It is important that 
the $\alpha$ breaking significantly affects 
not only the low-lying states but also
the excited $3\alpha$ cluster states.    
Firstly, the experimental large level spacing between the $0^+_1$ and 
$2^+_1$ states is described by the energy gain of the spin-orbit force 
due to the $\alpha$ breaking in 
the $0^+_1$, which contains the $p_{3/2}$-shell closure component.
It was found that 
the $\alpha$ breaking component 
is slightly contained even in the excited states with the developed $3\alpha$ 
cluster structure through the mixing of the ground state wave functions
with the $p_{3/2}$-shell closure component.
As a result of the mixing of the $\alpha$ breaking component, 
the significant $\beta$ decay strengths
are well reproduced by the present wave functions.
Moreover, it is important that
the non-$3\alpha$ component in the ground state
may change the $3\alpha$ cluster 
structure in the excited $0^+$ states through the orthogonality to the $0^+_1$ state.

In the analysis of the intrinsic structure and the 
$3\alpha$ clustering in the present results, 
it was found that the features of the 
$3\alpha$ cluster components well correspond
to those of the $3\alpha$GCM calculations\cite{GCM}.
As well as in the $3\alpha$GCM, we obtained the $0^+_2$, $2^+_2$ and $0^+_3$ 
states. The $0^+_2$, which corresponds to the observed $0^+$(7.65 MeV),
is dominated by [$^8$Be$(0^+)\otimes l=0]_{J=0}$, 
and shows the trend of cluster-gas features.
The predicted $0^+_3$ can not be directly assigned to the experimental
broad $0^+$ resonance(10.3 MeV). We expect that the corresponding $0^+$ 
spectra might be overlapped by the broad $0^+$(10.3 MeV).
The predicted $2^+_2$ state 
should be assigned to the recently observed $2^+$ state at 
$E_x=9.9$ MeV\cite{itoh04}.
In the weak coupling picture of $^8$Be+$\alpha$,
This $2^+_2$ can be regarded as the rotational member 
built on the $0^+_2$ state,  
because this state is dominated by [$^8$Be$(0^+)\otimes l=2]_{J=2}$.
We also proposed an alternative interpretation for the band assignment of the
the $2^+_2$ states
based on the remarkable $B(E2;2^+_2\rightarrow 0^+_3)$ 
and the similarity of the intrinsic structure with the $0^+_3$ in the
present results.
Namely, in a collective picture, the $0^+_3$ and the $2^+_2$ states 
are considered to be the members of 
the rotational band, which is formed by the deformed intrinsic state
with the linear-like $3\alpha$ structure.

In the present work, the number of the 
base wave functions is limited and the $3\alpha$ configurations 
with the large distance are not incorporated enough.  
Therefore, the description of the detailed resonant behavior is 
insufficient in the present framework.
Especially, for the broad resonances with $3\alpha$ clustering, 
the boundary conditions should be carefully treated in the calculation 
of the excitation energies and the widths. 
The broad $0^+$ and $2^+$ resonances have been theoretically studied
with the CSM and ACCC methods in the $3\alpha$ cluster models.
Such methods are promising in further study on these resonances.

In conclusion, we should stress that 
the cluster and the shell-model-like features coexist in 
$^{12}$C. 
Since these two kinds of nature interplay with each other,
one should take into account both the features appropriately 
for a systematic study of $^{12}$C. 
We expect that it is also significant for clarification of
the properties of the $3\alpha$ resonances.

\acknowledgments

The author would like to thank Prof. H. Horiuchi, Prof. A. Tohsaki, 
Prof. P. Schuck, Dr. C. Kurokawa and Dr. Y. Funaki 
for many discussions. She is  also thankful to 
members of Yukawa Institute for Theoretical Physics(YITP) 
in Kyoto University for valuable discussions.
The computational calculations in this work were supported by the 
Supercomputer Projects 
of High Energy Accelerator Research Organization(KEK)
and also the super computers of YITP.
This work was supported by Japan Society for the Promotion of 
Science and a Grant-in-Aid for Scientific Research of the Japan
Ministry of Education, Science and Culture.
The work was partially performed in the ``Research Project for Study of
Unstable Nuclei from Nuclear Cluster Aspects'' sponsored by
Institute of Physical and Chemical Research (RIKEN).


\section*{References}
\begin{thebibliography}{9}

\bibitem{SHELL1}
A. G. M. van Hees and P. W. M. Glaudemans, Z. Phys. A {\bf 315}, 223 (1984).
\bibitem{SHELL2}
A. A. Wolters, A. G. M. van Hees, and P. W. M. Glaudemans,
Phys. Rev. C {\bf 42}, 2053 (1990).
\bibitem{navratil03}
P. Navratil and W. E. Ormand,
Phys. Rev. C {\bf 68}, 034305 (2003).
\bibitem{RGM}
Y. Fukushima and M. Kamimura,
{\it Proc. Int. Conf. on Nuclear Structure, Tokyo, 1977,
edited by T. Marumori}[J. Phys. Soc. Jpn. {\bf 44}, 225 (1978);
M. Kamimura, Nucl. Phys. {\bf A351}, 456 (1981).
\bibitem{OCM}
H. Horiuchi, Prog. Theor. Phys. {\bf 51}, 1266 (1974); {\bf 53}, 447 (1975)
\bibitem{GCM}
E. Uegaki, S. Okabe, Y. Abe and H. Tanaka, Prog. Theor. Phys. {\bf 57},
1262 (1977).
E. Uegaki, Y. Abe, S. Okabe and H. Tanaka, Prog. Theor. Phys. {\bf 59},
 1031 (1978); {\bf 62}, 1621 (1979).
\bibitem{Fujiwara80}
Y. Fujiwara {\em et al.}, Prog. Theor. Phys. Suppl.{\bf 68}, 29 (1980).
\bibitem{descouvemont87}
P. Descouvemont and D. Baye, Phys. Rev. C {\bf 36}, 54 (1987).
\bibitem{Tohsaki01}
A. Tohsaki, H. Horiuchi, P. Schuck, and G. R\"opke, 
Phys. Rev. Lett. {\bf 87}, 192501 (2001).
\bibitem{Ropke98}
G. R\"opke, A. Schnell, P. Schuck, and P. Nozieres, Phys. Rev. Lett. {\bf 80},
3177 (1998).
\bibitem{john03}
B. John {\em et al.}, Phys. Rev. C {\bf 68}, 14305 (2003).
\bibitem{fynbo03}
H. O. U. Fynbo {\em et al.}, Phys. Rev. Lett. {\bf 91}, 082502 (2003);
H. O. U. Fynbo {\em et al.}, Nucl. Phys. {\bf A738}, 59 (2004).
\bibitem{itoh04}
M. Itoh {\em et al.}, Nucl. Phys. {\bf A738}, 268 (2004).
\bibitem{diget05}
C. Aa. Diget {\em et al.}, Nucl. Phys. {\bf A760}, 3 (2005).
\bibitem{kurokawa04}
C. Kurokawa and K. Kato,
Nucl. Phys. {\bf A738}, 455 (2004).
\bibitem{kurokawa05}
C. Kurokawa and K. Kato,
Phys. Rev. C {\bf 71}, 021301(R) (2005).
\bibitem{funaki05}
Y. Funaki, A. Tohsaki, H. Horiuchi, P. Schuck, and G. R\"opke,
Eur. Phys. J. {\bf A24}, 321(2005).
\bibitem{funaki06}
Y. Funaki, H. Horiuch and A. Tohsaki,
Prog. Theor. Phys. {\bf 115}, 115 (2006).
\bibitem{ENYO-c12}
 Y. Kanada-En'yo,
Phys. Rev. Lett. {\bf 81}, 5291 (1998).
\bibitem{ENYO-be10}
Y. Kanada-En'yo, H. Horiuchi and A. Dot\'{e}, Phys. Rev. C {\bf 60}, 
064304 (1999).
\bibitem{ENYO-be12}
Y. Kanada-En'yo, H. Horiuchi
Phys. Rev. C {\bf 68}, 014319 (2003).
\bibitem{Itagaki04}
N. Itagaki, S. Aoyama, S. Okabe and K. Ikeda,
Phys. Rev. C {\bf 70}, 054307 (2004).
\bibitem{Neff-c12}
T. Neff and H. Feldmeier, Nucl. Phys. {\bf A738}, 357 (2004).
\bibitem{ENYObc}
 Y. Kanada-En'yo, H. Horiuchi and A. Ono,
Phys. Rev. C {\bf 52}, 628 (1995);
 Y. Kanada-En'yo and H. Horiuchi,
Phys. Rev. C {\bf 52}, 647 (1995).
\bibitem{ENYOsup}
Y. Kanada-En'yo and  H. Horiuchi, Prog. Theor. Phys. Suppl.{\bf 142},
 205 (2001).
\bibitem{AMDrev}
Y. Kanada-En'yo, M. Kimura and H. Horiuchi, Comptes rendus Physique Vol.4, 
497 (2003).
\bibitem{TOHSAKI}
 T. Ando, K.Ikeda, and A. Tohsaki, Prog. Theor. Phys.
 {\bf 64}, 1608 (1980).
\bibitem{LS}
 N. Yamaguchi, T. Kasahara, S. Nagata, and Y. Akaishi,
 Prog. Theor. Phys. {\bf 62}, 1018 (1979);
 R. Tamagaki, Prog. Theor. Phys. {\bf 39}, 91 (1968).
\bibitem{nucldata}
H. DE Vries {\em et al.}, 
Atomic Data and Nuclear Data Tables, {\bf 36} (1987). 
\bibitem{funaki-nucl06}
Y. Funaki, A. Tohsaki, H. Horiuchi, P. Schuck, and G. R\"opke, 
nucl-th/0601035.
\bibitem{eyrich87}
W. Eyrich {\em et al.}, Phys. Rev. C {\bf 36}, 416 (1987). 
\bibitem{isotopes}
F. Ajzenberg-Selove and J.H. Kelley, Nucl. Phys. {\bf A506}, 1 (1990).
\bibitem{Sick70}
I. Sick and J. S. McCarthy, Nucl. Phys. {\bf A150}, 631 (1970); 
A. Nakada, Y. Torizuka and Y. Horikawa, Phys. Rev. Lett. {\bf 27}, 745 (1971);
and 1102 (Erratum); P. Strehl and Th. H. Schucan, Phys. Lett. {\bf 27B}, 
641 (1968).
\bibitem{morinaga66}
H. Morinaga, Phys. Rev. {\bf 101}, 1956 (1956);
Phys. Lett. {\bf 21}, 78(1966).
\bibitem{suzuki72}
Y. Suzuki, H. Horiuchi and K. Ikeda, Prog. Theor. Phys. {\bf 47}, 
1517 (1972).   
\bibitem{funaki03}
Y. Funaki, A. Tohsaki, H. Horiuchi, P. Schuck, and G. R\"opke,
Phys. Rev. C {\bf 67}, 051306(R) (2003).
\bibitem{Pichler97}
R. Pichler, H. Oberhummer, A. Cs\'ot\'o and S. A. Moszkowski, Nucl. Phys. 
{\bf A618}, 55 (1997).




\end{thebibliography}
\end{document}